\documentclass[10pt,journal,doublecolumn,twoside]{IEEEtran}

\ifCLASSOPTIONcompsoc
\usepackage[nocompress]{cite}
\else
\usepackage{cite}
\fi
\usepackage{makecell}
\usepackage{tabularx}
\usepackage{array}
\usepackage{graphicx}
\usepackage{amsmath,amssymb,amsfonts,bm}
\usepackage{algorithm}
\usepackage{algorithmic}
\usepackage{array,booktabs,multirow,tabularx}
\usepackage{stfloats}
\usepackage{placeins}
\usepackage{url}
\usepackage{xcolor}
\usepackage{siunitx}
\usepackage{enumitem}
\usepackage{etoolbox}
\usepackage{hyperref}
\hypersetup{hidelinks,colorlinks=true,allcolors=black,pdfstartview=Fit,breaklinks=true}

\newcommand{\mathbbm}[1]{\mathbb{#1}}

\graphicspath{{figure/}{figures/}{./}}
\begin{document}

\title{Multi-Granularity Communication-Computing Resource Mapping for User Association in MEO-Assisted LEO Satellite Networks}
\author{Yi~Jing, \emph{Graduate Student Member, IEEE,}
	Chunxiao Jiang, \emph{Fellow, IEEE}, Jiawei Wang, \emph{Member, IEEE}, \\Jiachen Sun, \emph{Graduate Student Member, IEEE,} Yafeng Zhan, \emph{Member, IEEE}
	\thanks{This work was supported by the National Natural Science Foundation of China under Grant 62341131, Grant 62325108, Grant 62501354 and Grant 62131012. \textit{(Corresponding author: Chunxiao Jiang.)}}
	\thanks{Y. Jing and J. Sun are with the Department of Electronic Engineering, Tsinghua University, Beijing, 100084, China. (e-mail: jy21@mails.tsinghua.edu.cn, sjc20@tsinghua.org.cn.)}
	\thanks{C. Jiang, J. Wang and Y. Zhan are with the State Key Laboratory of Space Network and Communications, Tsinghua University, Beijing 100084, China, and also with the Beijing National Research Center for Information Science and Technology, Tsinghua University, Beijing 100084, China (e-mail: jchx@tsinghua.edu.cn, w-jw@tsinghua.edu.cn, zhanyf@tsinghua.edu.cn.)}
}

\markboth{IEEE Transactions on Network Science and Engineering}
{Jing \MakeLowercase{\textit{et al.}}: Resource Mapping for User Association in MEO-Assisted LEO Networks}

\IEEEtitleabstractindextext{
\begin{abstract}
Low-Earth-orbit (LEO) satellite networks are evolving into spaceborne communication--computing infrastructures. Their user association must jointly account for the radio link and the CPU, GPU, NPU, memory, cache, storage, energy, queue, and thermal states that determine executable service capacity. This paper studies a MEO-assisted LEO architecture in which medium-Earth-orbit (MEO) regional state aggregators collect and relay time-stamped telemetry, the network operation control center (NOCC) maintains the global logical resource view and makes the final association decisions, and LEO satellites provide access and onboard execution. We develop a multi-granularity representation and service-primitive-driven mapping that converts a service-level request into a target-LEO-dependent quantized resource vector. The integer formulation, exact mixed-integer linear programming (MILP) reference, and capacity-aware primal--dual algorithm share the same mapped objective. Experiments use real-world service bandwidth traces, a 1584-satellite LEO shell, multiple traffic seeds, controlled load and mapping error, delayed telemetry, and temporal sessions. Relative to static mapping, the proposed method improves realized utility by 10.64\% and lowers end-to-end outage by 5.70 percentage points. Target quantization reduces signaling by 19.84\% with negligible utility loss. Additional load, delay, temporal, runtime, and exact-MILP audits delimit the operating region and remaining mapping uncertainty.

\end{abstract}

}

\maketitle
\IEEEdisplaynontitleabstractindextext
\IEEEpeerreviewmaketitle

\section{Introduction}
\label{sec:introduction}

Low-Earth-orbit (LEO) satellite networks are expanding from transparent forwarding and broadband access to onboard computing, caching, and intelligent service execution~\cite{9210567,9861699,9711564,10209551,10409745,10886927,10937133,9964003,9613260,11063433,9344666,9632811,11744392}. This evolution is important for non-terrestrial networks, emergency access, maritime and remote-area services, low-altitude aerial users, and AI-assisted applications. The request observed by the network may contain bandwidth, latency, reliability, priority, service type, quality level, cache dependency, and hidden computing demand. User association must therefore be revisited from a communication--computing resource perspective.

A conventional LEO user association strategy selects the visible satellite by SINR, elevation, distance, or communication load~\cite{10533435,10645749,9870798,11244216,10947167,10938640,10892320}. This becomes insufficient for onboard execution: video forwarding mainly consumes bandwidth and buffering, whereas video analytics additionally requires decoding, GPU/NPU inference, memory, cache, and energy. A language-model query may likewise have little input traffic but high accelerator and memory demand. Executable service capacity therefore depends on both the radio link and onboard resources.

Mapping-aware association infers the target-dependent demand before selecting a LEO. Section~\ref{sec:results} specifies the resource profiles and service settings.

This paper focuses on a MEO-assisted LEO user-association architecture motivated by visibility-aware association, satellite beam/resource scheduling, LEO topology management, and recent LEO--MEO sensing and forecasting studies~\cite{11288045,10582841,11091526,11489000,11524084,11755923}. MEO satellites act as regional state aggregators: they collect, time-stamp, deduplicate, and relay LEO summaries through the ground segment. The NOCC is the single logical decision authority, fuses the regional reports into a constellation-wide view, executes mapping and association, and returns compact commands. LEO satellites remain the service and data plane. This hierarchy provides regional collection paths and a global optimization scope, while delayed or missing reports expose the controller to state staleness. The centralized logical authority also introduces scalability and availability requirements; replicated NOCC instances, hot standby, and distributed database consensus are practical deployment mechanisms, while the present model evaluates one logically centralized instance. Future extensions will incorporate inter-LEO computation offloading and MEO execution.

Three challenges arise. First, LEO resources are heterogeneous in granularity. Communication resources are organized into spectrum blocks, beam-time units, rate levels, queues, and handover overheads. Computing resources include CPU, GPU, NPU, memory, cache, storage, and energy, whose availability and scheduling granularity are highly different. Second, there is a request-resource mapping gap. Existing joint communication--computing optimization often assumes that task size, CPU cycles per bit, and deadlines are already known. In practice, the control layer usually observes only service-level requirements, while the concrete demand for compute, memory, storage, cache, and energy must be inferred before association. Third, capacity-oriented association must enforce multiple resource constraints while maximizing successfully executable service value; outage, mapping error, waste, and signaling quantify efficiency and robustness.

To address these challenges, we develop a model-driven association framework consisting of multi-granularity resource representation (MGRR), service-primitive-driven communication--computing resource mapping (SP-CCRM), target-LEO quantized resource mapping (TL-QRM), capacity-aware primal--dual user association (CAP-PD-UA), and online mapping calibration (OMC). The main contributions are summarized as follows.
\begin{itemize}[leftmargin=*]
	\item We introduce MGRR together with SP-CCRM and TL-QRM as a target-dependent mapping interface. Physical, scheduling, and reporting granularities describe communication, compute, memory, cache, storage, and energy resources, while service primitives convert observable requests into candidate-specific quantized demands and explicitly account for reconfiguration and allocation waste.
	\item We formulate mapping-aware association with one objective shared by the online optimizer and exact MILP reference. Pairwise compatibility is separated from aggregate capacity, and CAP-PD-UA converts multidimensional scarcity into resource prices followed by finite primal recovery, thereby returning a capacity-feasible assignment even when the discrete price iteration reaches its limit.
	\item We evaluate mapping, quantization, association, delayed control, and temporal sessions under trace-driven traffic and a propagated 1584-satellite shell. Multi-seed paired tests, common-load and common-error comparisons, structured ablations, serial runtime measurements, and exact-MILP audits identify the source, persistence, and operating limits of the performance gain.
\end{itemize}

The remainder of this paper is organized as follows. Section~\ref{sec:related} reviews related work. Section~\ref{sec:system} introduces the system model. Section~\ref{sec:problem} formulates the mapping and association problem. Section~\ref{sec:scheme} presents the proposed scheme. Section~\ref{sec:results} reports simulation results. Section~\ref{sec:conclusion} concludes the paper.

\section{Related Work}
\label{sec:related}

We review three closely related lines: LEO association and satellite edge computing, service orchestration and heterogeneous resources, and communication--computing service metrics.

\subsection{LEO Edge Association}

LEO association studies select serving satellites by received power, SINR, elevation, distance, load, or handover stability~\cite{10902219,11131136,11003898}. Recent work extends this scope to integrated satellite--HAPS--ground access, learning-based handover, SAGIN slicing, and computation offloading~\cite{10304301,10371217,10304325,10502326}. These methods improve access and mobility while leaving request-level execution feasibility implicit; a strong-link LEO may still lack accelerator, memory, cache, energy, or thermal headroom.

Satellite edge computing addresses offloading, caching, slicing, learning-based scheduling, and joint communication--computation allocation~\cite{10227270,10715709,10916814,10443227,10566891,10589437,10478101,10147866,10061467,11627876}. Further studies consider industrial IoT, GEO--LEO cooperation, distributed execution, DNN partitioning, and federated learning~\cite{10042025,10522487,10107794,10829804,11173862,10716798,10689514,10198334,10032271,11066269}. Most assume known task size, CPU intensity, and deadline; our mapping constructs target-dependent demands from the service type and QoS fields observed at access.

\subsection{Heterogeneous Orchestration}

Service-chain and SAGIN orchestration optimize placement, routing, scheduling, migration, or slicing across satellites, terrestrial edges, and clouds~\cite{10856408,10416394,10323351,10207691,10738455,10263617}. Application-specific state also enters physical-layer objectives, as in sensing-assisted secure beamforming with leakage suppression~\cite{zhao2026sensing}. These works generally specify resource demands before optimization. Our primitive decomposition constructs a candidate-local resource vector at access, with cross-LEO placement and migration reserved for future work.

Heterogeneous satellite-resource studies consider CPU allocation, energy-aware offloading, orbital computing, cloud--edge integration, scheduling, fairness, and virtual embedding~\cite{10380438,9894082,10143230,10758345,10714036,10414013,11755775}. Aggregate CPU rate or AI throughput cannot jointly express GPU/NPU compatibility, slicing, VRAM/SRAM, model cache, scheduling quanta, reconfiguration, and thermal limits. MGRR retains these distinctions across physical, scheduling, and reporting granularities.

\subsection{Research Positioning}

Satellite and SAGIN systems use throughput, outage, delay, energy, AoI, signaling, reliability, and allocation metrics~\cite{10440193,9978924,10032237,11153682,9167267,9874801}. Link throughput or admission count can overestimate capability when onboard execution fails; our service-capacity metric requires successful association and satisfaction of all communication--computing constraints.

The framework connects multi-granularity representation, primitive-driven mapping, and target-aware association. Mapping ablations isolate request-to-resource inference; exact-MILP and runtime comparisons assess association quality and deployability.

\FloatBarrier
\section{System Model}
\label{sec:system}

\begin{figure}[!t]
	\centering
	\includegraphics[width=\linewidth]{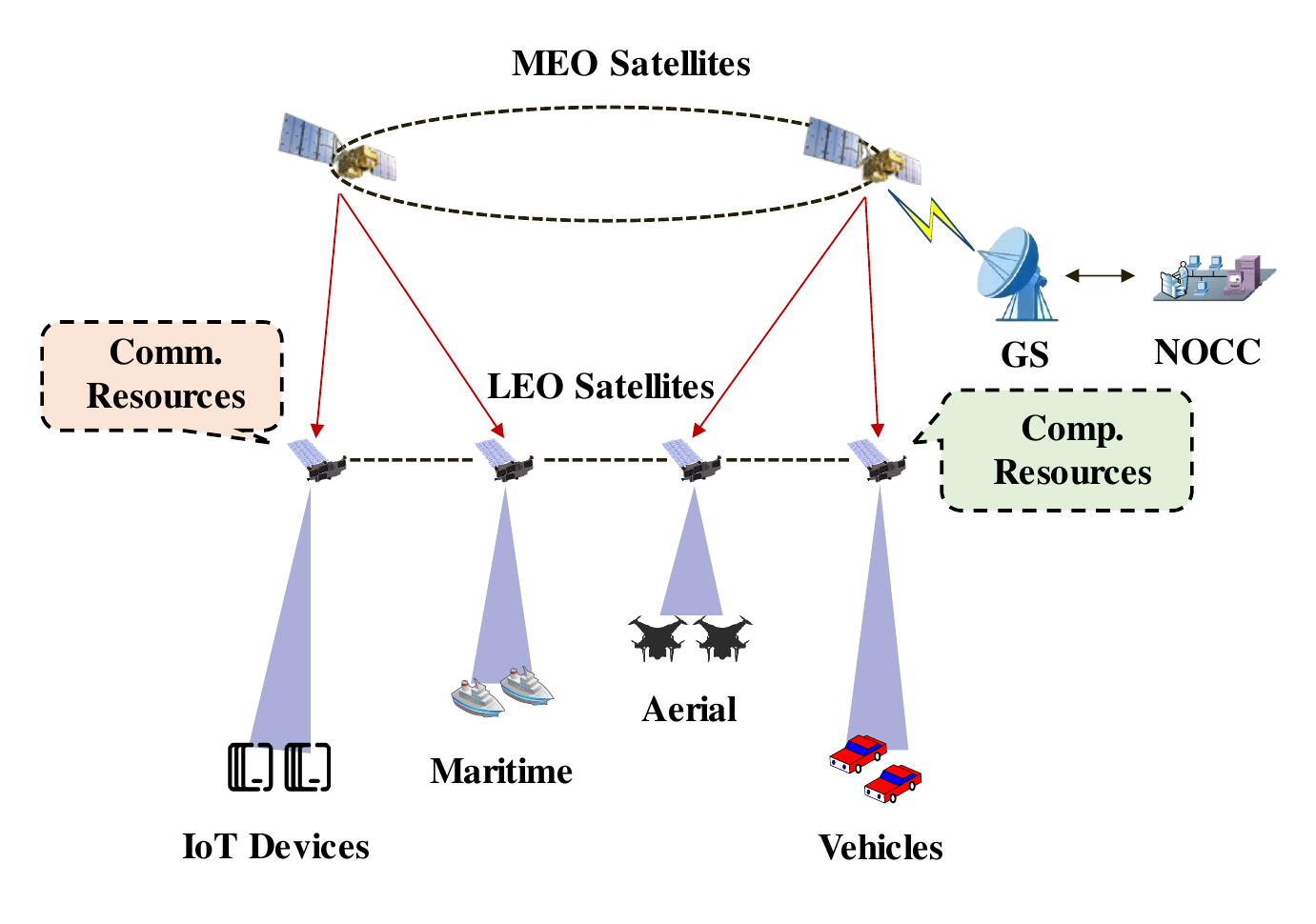}
	\caption{Logical system architecture. MEO regional state aggregators relay time-stamped LEO telemetry through the ground segment; the NOCC fuses a global resource view and issues association commands; LEO satellites provide access and onboard execution.}
	\label{fig:system_model}
\end{figure}

\subsection{Network Model}

As shown in Fig.~\ref{fig:system_model}, we consider a slotted MEO-assisted LEO network that supports heterogeneous users, including IoT devices, maritime users, aerial users, vehicles, and broadband or emergency terminals. The network follows the trend of collaborative satellite--terrestrial edge computing, service-aware LEO--terrestrial orchestration, microservice deployment in space computing power networks, cache-enabled hybrid satellite--aerial--terrestrial systems, and distributed satellite-edge service deployment~\cite{9999289,9520123,10713242}. The architecture has four roles. The user layer generates service requests. The LEO layer provides data-plane access and onboard execution under finite multidimensional capacities. MEO satellites serve as regional state aggregators and telemetry relays. Ground gateways carry reports and commands between the space segment and the NOCC. The NOCC maintains the single logical global state and is the sole final authority for TL-QRM and CAP-PD-UA decisions.

The user set, LEO-satellite set, MEO-aggregator set, and resource-type set in slot $t$ are denoted by $\mathcal U_t$, $\mathcal S$, $\mathcal M$, and $\mathcal R$, respectively. Semantic resource labels distinguish communication, computing, and storage resources:
\begin{equation}
\begin{aligned}
\mathcal R=\{&\mathrm{bw},\mathrm{cpu},\mathrm{gpu},\mathrm{npu},\mathrm{vram},\\
&\mathrm{sram},\mathrm{cache},\mathrm{sto},\mathrm{ene}\},
\end{aligned}
\label{eq:resource_set}
\end{equation}
where $\mathrm{bw}$, $\mathrm{cpu}$, $\mathrm{gpu}$, $\mathrm{npu}$, $\mathrm{vram}$, $\mathrm{sram}$, $\mathrm{cache}$, $\mathrm{sto}$, and $\mathrm{ene}$ denote communication bandwidth, CPU, GPU, NPU, video memory, SRAM, cache, storage, and energy, respectively. Here, $\mathrm{bw}$ represents the radio and link-side service capability, while the other resources represent onboard service-execution capability. Different users in Fig.~\ref{fig:system_model} may consume different combinations of these resources. For example, IoT services mainly stress access scheduling and lightweight processing, whereas AI-assisted aerial or emergency services may simultaneously require stable links, accelerator resources, memory/cache, storage, and energy.

The binary association variable is
\begin{equation}
	x_{u,i,t}=\begin{cases}
		1, & \text{if user }u\text{ accesses LEO }i\text{ in slot }t,\\
		0, & \text{otherwise}.
	\end{cases}
	\label{eq:association_var}
\end{equation}
Each admitted user is associated with at most one LEO satellite:
\begin{equation}
	\sum_{i\in\mathcal S} x_{u,i,t}\le 1,\quad \forall u\in\mathcal U_t.
	\label{eq:unique_assoc}
\end{equation}

The reporting domain of aggregator $m$ evolves with orbital visibility. Let $\mathcal S_m^{\rm vis}(t)=\{i\in\mathcal S:L_{m,i}(t)=1\}$ denote the LEOs whose reports can reach $m$ in slot $t$. These visibility sets may overlap. A primary reporting aggregator
\begin{equation}
	m_i^\star(t)=\arg\max_{m:i\in\mathcal S_m^{\rm vis}(t)}\upsilon_{m,i}^{\rm ctrl}(t)
	\label{eq:primary_aggregator}
\end{equation}
is selected from the feasible control links using control-link quality $\upsilon_{m,i}^{\rm ctrl}(t)$. The corresponding primary sets form a partition for accounting, while overlapping visibility supplies redundant reports. During a primary change, the old and new aggregators may both forward one reporting epoch. Reports carry the tuple $(i,t_i^{\rm stamp},n_i^{\rm ver})$; the NOCC removes duplicate versions and retains the newest valid report. This procedure gives each LEO one authoritative record in the fused global view while preserving redundant reception.

Telemetry and commands have finite delay. Let $\zeta_{i,t}^{\rm age}$ be the age, in slots, of the newest report from LEO $i$ available when slot $t$ is scheduled. MEO regional state aggregator $m\in\mathcal M$ constructs the delayed view
\begin{equation}
	\begin{aligned}
		\hat{\bm z}_{m,t}=\{&\hat C_{i,r,t-\zeta_{i,t}^{\rm age}},\hat\theta_{u,i,t},
		\hat Q_{i,t-\zeta_{i,t}^{\rm age}}^{\rm que},\hat{\bm\sigma}_{i,t-\zeta_{i,t}^{\rm age}}^{\rm acc},\\
		&\hat h_{i,t-\zeta_{i,t}^{\rm age}}^{\rm cache},
		\hat H_{i,t-\zeta_{i,t}^{\rm age}}^{\rm th}\}_{i\in\mathcal S_m^{\rm vis}(t),r\in\mathcal R},
	\end{aligned}
	\label{eq:meo_view}
\end{equation}
where a hat denotes the controller-side estimate, $\bm\sigma_i^{\rm acc}$ is the accelerator-availability vector, $h_i^{\rm cache}$ is the cache state, and $H_i^{\rm th}$ is thermal headroom. The NOCC fuses $\{\hat{\bm z}_{m,t}\}_{m\in\mathcal M}$ after version checking and deduplication into $\hat{\bm z}^{\rm G}_t$ over the complete set $\mathcal S$. Consequently, candidates observed through different MEO regions are compared in the same global optimization. At the beginning of a decision epoch, the reported queue, accelerator, cache, and thermal states are frozen while candidate mappings are evaluated. Assignment-induced bandwidth, computation, memory, energy, and temperature increments remain decision variables and are enforced by the aggregate constraints. Geometry and link-quality estimates are held during one decision slot and refreshed at the next boundary. Within-slot variation appears as execution-side mapping/QoS error.

The NOCC runs TL-QRM and CAP-PD-UA using $\hat{\bm z}^{\rm G}_t$, obtains $x_{u,i,t}$, and sends commands through an available ground--MEO--LEO control path. LEO satellites execute access, transmission, and onboard computing and return measured consumption and service states for calibration. MEO satellites aggregate and relay state; LEO satellites execute user tasks. The numerical model represents the NOCC as one logical controller. Replicated databases and hot-standby NOCC instances can improve availability in deployment. The experiments cover controller delay, report age, candidate pruning, and serial runtime; future work will address correlated control-plane failures and consensus among replicated NOCC instances.
\subsection{User Model}
A user request is represented as
\begin{equation}
\begin{aligned}
    \bm\varrho_{u,t}=(&\tau_u,\psi_u,B_u^{\rm ul},B_u^{\rm dl},T_u^{\max},\\
    &d_u,Q_u,p_u,\eta_u),
\end{aligned}
\label{eq:user_request}
\end{equation}
where $\tau_u$ is the user type, $\psi_u$ is the service type, $B_u^{\rm ul}$ and $B_u^{\rm dl}$ are uplink and downlink bandwidth requests in Mbit/s, $T_u^{\max}$ is the latency bound, $d_u$ is the requested number of service slots, $Q_u$ is the quality parameter, $p_u$ is the priority, and $\eta_u$ is the reliability requirement. A session admitted at $t_u^0$ remains active until completion, drop, or visibility loss, and its held resource in slot $t$ is included in the same aggregate capacity constraint as a new request. If the serving satellite changes, $T^{\rm ho}_{u,i,t}$ and the corresponding state-transfer demand are charged in that slot. The single-slot formulation below is consequently the per-epoch subproblem of this temporal process; the temporal experiment evaluates persistence, blocking, drops, and handovers explicitly.

\subsection{Visibility Model}
Let $\theta_{u,i,t}$ denote the elevation angle from user $u$ to LEO $i$. The visibility indicator is
\begin{equation}
    v_{u,i,t}=\mathbbm 1\{\theta_{u,i,t}\ge \theta^{\min}\},
    \label{eq:visibility}
\end{equation}
where $\theta^{\min}$ is the minimum elevation angle. The candidate LEO set is
\begin{equation}
    \mathcal S_{u,t}^{\rm cand}=\{i\in\mathcal S\mid v_{u,i,t}=1\}.
    \label{eq:candidate_set}
\end{equation}
The visibility constraint is
\begin{equation}
    x_{u,i,t}\le v_{u,i,t},\quad \forall u,i,t.
    \label{eq:visibility_cons}
\end{equation}

\subsection{Service Model}
The service-primitive set is
\begin{equation}
\begin{aligned}
\mathcal P=\{&\text{access},\text{transmission},\text{security},\text{protocol},\\
&\text{preprocess},\text{inference},\text{cache},\text{output}\}.
\end{aligned}
\label{eq:primitive_set}
\end{equation}
For service type $\psi_u$, the primitive chain is
\begin{equation}
    \mathcal C(\psi_u)=(p_{u,1},p_{u,2},\ldots,p_{u,L_u}).
    \label{eq:primitive_chain}
\end{equation}
The input data volume in one slot is
\begin{equation}
    D_{u,t}^{(0)}=10^6\Delta_t B_u^{\rm ul},
    \label{eq:data_input}
\end{equation}
where $\Delta_t$ is the slot duration and $D_{u,t}^{(0)}$ is measured in bits. After primitive $\ell$, the data volume evolves as
\begin{equation}
    D_{u,t}^{(\ell)}=\gamma_{p_{u,\ell}}D_{u,t}^{(\ell-1)},
    \quad \ell=1,\ldots,L_u,
    \label{eq:data_evolution}
\end{equation}
where $\gamma_{p}$ is the output-to-input data ratio of primitive $p$. For $p=p_{u,\ell}$, the input processed by that primitive is denoted by $D_{u,t}^{p}=D_{u,t}^{(\ell-1)}$. A primitive that preserves data volume has $\gamma_p=1$; preprocessing and inference use the ratios in Table~\ref{tab:primitive_profiles}. The final output volume is $D_{u,t}^{\rm out}=D_{u,t}^{(L_u)}$.

\subsection{Multi-Granularity Resource Model}
The resource atom of LEO $i$ for resource $r\in\mathcal R$ is defined as
\begin{equation}
\begin{aligned}
    \mathcal A_{i,r,t}=(&C_{i,r,t},\delta_{i,r}^{\rm phy},\delta_{i,r}^{\rm sch},
    \delta_{i,r}^{\rm rep},\\
    &\omega_{i,r},h_{i,r},\mu_{i,r},\xi_{i,r}),
\end{aligned}
\label{eq:resource_atom}
\end{equation}
where $C_{i,r,t}$ is nominal remaining capacity; $\delta_{i,r}^{\rm phy}$ is the hardware resolution (e.g., a core, accelerator slice, or memory bank); $\delta_{i,r}^{\rm sch}$ is the allocation quantum used by TL-QRM; and $\delta_{i,r}^{\rm rep}$ is the telemetry quantum. Parameters $\omega_{i,r}$, $h_{i,r}$, $\mu_{i,r}$, and $\xi_{i,r}$ denote reconfiguration overhead, minimum holding time, management reserve, and effective efficiency, respectively. The effective remaining capacity is
\begin{equation}
    \bar C_{i,r,t}=\xi_{i,r}\left(C_{i,r,t}-\mu_{i,r}\right).
    \label{eq:effective_capacity}
\end{equation}
The operational roles of physical and reporting granularity are
\begin{equation}
 C_{i,r,t}=n_{i,r,t}^{\rm phy}\delta_{i,r}^{\rm phy},\qquad
 \hat C_{i,r,t}=\left\lfloor\frac{\widetilde C_{i,r,t}}{\delta_{i,r}^{\rm rep}}\right\rfloor\delta_{i,r}^{\rm rep},
 \label{eq:physical_reporting_quantization}
\end{equation}
where $n_{i,r,t}^{\rm phy}$ is the available number of physical atoms and $\widetilde C_{i,r,t}$ is the delayed controller estimate of $\bar C_{i,r,t}$, clipped to the physical range. The reported value $\hat C_{i,r,t}$ is the capacity available to the NOCC optimization. A continuous request first occupies physical atoms,
\begin{equation}
q_{u,i,t}^{r,\rm phy}=
\left\lceil\frac{q_{u,i,t}^{r}}{\delta_{i,r}^{\rm phy}}\right\rceil
\delta_{i,r}^{\rm phy},
\label{eq:physical_demand_quantization}
\end{equation}
and is then converted into scheduling quanta by~\eqref{eq:generic_quantization}.
This definition captures the discrete and overhead-bearing consumption of LEO resources. The three granularities can differ: a physical accelerator can expose several scheduling slices, while a report may encode only a coarse utilization bin. Their separate effects are evaluated through the no-quantization/coarse-mapping ablations and signaling measurements.

\subsection{Communication Model}
For a candidate pair $(u,i)$, the end-to-end communication time is
\begin{equation}
\begin{aligned}
T_{u,i,t}^{\rm com}=&T_{u,i,t}^{\rm ho}+T_{u,i,t}^{\rm tx}+T_{u,i,t}^{\rm prop}\\
&+T_{u,i,t}^{\rm q}+T_{u,i,t}^{\rm proc}.
\end{aligned}
\label{eq:comm_delay}
\end{equation}
Here, $T^{\rm ho}$ is handover/configuration overhead, $T^{\rm tx}$ is transmission time, $T^{\rm prop}$ is propagation time, $T^{\rm q}$ is communication-queueing time, and $T^{\rm proc}$ is communication processing time. The transmission time is
\begin{equation}
    T_{u,i,t}^{\rm tx}=\frac{D_{u,t}^{(0)}+D_{u,t}^{\rm out}}
    {10^6R_{u,i,t}},
    \label{eq:tx_delay}
\end{equation}
where $R_{u,i,t}$ is measured in Mbit/s and follows discrete rate levels:
\begin{equation}
    R_{u,i,t}\in\mathcal L_i=\{R_{i,1},R_{i,2},\ldots,R_{i,K_i}\}.
    \label{eq:rate_levels}
\end{equation}
The propagation term covers the input and result paths, $T_{u,i,t}^{\rm prop}=2d_{u,i,t}^{\rm slant}/c$, where $c$ is the speed of light. The queue term is evaluated from the newest measured backlog and service rate, $T_{u,i,t}^{\rm q}=\hat Q_{i,t-\zeta_{i,t}^{\rm age}}^{\rm que}/(\hat R_{i,t-\zeta_{i,t}^{\rm age}}^{\rm srv}+\varepsilon)$, where $\varepsilon>0$ is a numerical safeguard, and is held fixed during candidate evaluation. Newly admitted users are coupled through the aggregate resource constraints, while the within-slot queue estimate remains fixed. Execution-side perturbation and calibration capture the resulting approximation error. The processing delay is approximated as
\begin{equation}
    T_{u,i,t}^{\rm proc}=c_i^{\rm proc}D_{u,t}^{(0)}.
    \label{eq:proc_delay}
\end{equation}
where $c_i^{\rm proc}$ is the communication-processing time per input bit at LEO $i$.
Let $B_u^{\rm req}=\max\{B_u^{\rm ul},B_u^{\rm dl}\}$ and set the continuous bandwidth demand to $q_{u,i,t}^{\rm bw}=B_u^{\rm req}$. The number of bandwidth scheduling blocks is
\begin{equation}
    n_{u,i,t}^{\rm bw}=
    \left\lceil \frac{q_{u,i,t}^{\rm bw,phy}}
    {\delta_{i,\mathrm{bw}}^{\rm sch}}\right\rceil,
    \label{eq:comm_blocks}
\end{equation}
which yields actual allocated communication capacity
\begin{equation}
    \hat q_{u,i,t}^{\rm bw}=n_{u,i,t}^{\rm bw}\delta_{i,\mathrm{bw}}^{\rm sch}.
    \label{eq:comm_alloc}
\end{equation}
The communication-block waste is
\begin{equation}
    w_{u,i,t}^{\rm bw}=\hat q_{u,i,t}^{\rm bw}
    -B_u^{\rm req}.
    \label{eq:comm_waste}
\end{equation}

\subsection{Accelerator Model}
The effective CPU computing capacity of LEO $i$ is
\begin{equation}
    F_{i,t}^{\rm cpu}=\xi_{i,\mathrm{cpu}}F_{i,t}^{\rm cpu,nom}.
    \label{eq:cpu_capacity}
\end{equation}
The service-intrinsic reference CPU workload of user $u$ is
\begin{equation}
\begin{aligned}
q_{u,t}^{\rm cpu,ref}=\sum_{p\in\mathcal C(\psi_u)}\big(&\kappa_{p,\rm bit}^{\rm cpu}D_{u,t}^{p}
+\kappa_{p,\rm pkt}^{\rm cpu}N_{u,t}^{\rm pkt}\\
&+\kappa_{p,\rm frm}^{\rm cpu}N_{u,t}^{\rm frm}
+\kappa_{p,\rm tok}^{\rm cpu}N_{u,t}^{\rm tok}
+\kappa_{p,\rm req}^{\rm cpu}\big),
\end{aligned}
\label{eq:cpu_demand}
\end{equation}
where $N_{u,t}^{\rm pkt}$, $N_{u,t}^{\rm frm}$, and $N_{u,t}^{\rm tok}$ are the packet, frame, and token counts derived reproducibly from the request. TL-QRM adds candidate-specific overhead to this reference workload and forms $q_{u,i,t}^{\rm cpu}$ before quantization. The CPU block number is
\begin{equation}
    n_{u,i,t}^{\rm cpu}=\left\lceil
    \frac{q_{u,i,t}^{\rm cpu,phy}}{\delta_{i,\mathrm{cpu}}^{\rm sch}}
    \right\rceil.
    \label{eq:cpu_blocks}
\end{equation}
The reference GPU and NPU workloads are
\begin{equation}
\begin{aligned}
q_{u,t}^{\rm gpu,ref}&=z_u^{\rm gpu}\sum_{p\in\mathcal C(\psi_u)}
\left(\kappa_{p,\rm frm}^{\rm gpu}N_{u,t}^{\rm frm}
+\kappa_{p,\rm tok}^{\rm gpu}N_{u,t}^{\rm tok}\right),\\
q_{u,t}^{\rm npu,ref}&=z_u^{\rm npu}\sum_{p\in\mathcal C(\psi_u)}
\left(\kappa_{p,\rm frm}^{\rm npu}N_{u,t}^{\rm frm}
+\kappa_{p,\rm tok}^{\rm npu}N_{u,t}^{\rm tok}\right),
\end{aligned}
\label{eq:gpu_npu_demand}
\end{equation}
where $z_u^{\rm gpu}$ and $z_u^{\rm npu}$ indicate whether the service chain admits the corresponding accelerator implementation. TL-QRM tests both reference workloads against candidate $i$ and retains the selected accelerator demand in $q_{u,i,t}^{\rm gpu}$ or $q_{u,i,t}^{\rm npu}$. The implemented accelerator-memory rules are
\begin{equation}
\begin{aligned}
q_{u,t}^{\rm vram,ref}={}&z_u^{\rm gpu}\left(M_u^{\rm model}
+0.10\frac{D_{u,t}^{(0)}}{8\times10^6}\right.\\
&\left.\qquad+M_u^{\rm batch}+64+M_u^{\rm kv}\right),\\
q_{u,t}^{\rm sram,ref}={}&z_u^{\rm npu}M_u^{\rm work},
\end{aligned}
\label{eq:accelerator_memory}
\end{equation}
in megabytes. Here $M_u^{\rm batch}=128$ for video/emergency AI and 64 for other accelerated services; $M_u^{\rm kv}$ is 1024 for LLM, 256 for small LLM, and zero otherwise. The NPU working set $M_u^{\rm work}$ is 96 for video/emergency AI, 64 for ASR/small LLM, 512 for LLM, and 32 otherwise. The model sizes $M_u^{\rm model}$ are listed in Table~\ref{tab:service_settings}. After accelerator selection, $q_{u,i,t}^{\rm vram}=q_{u,t}^{\rm vram,ref}$ for GPU execution and $q_{u,i,t}^{\rm sram}=q_{u,t}^{\rm sram,ref}$ for NPU execution; the unselected memory demand is zero. A GPU slice is represented by
\begin{equation}
    \mathcal G_{i,a}=(\Omega_{i,a}^{\rm gpu},M_{i,a}^{\rm vram},
    H_{i,a}^{\rm bw},\omega_{i,a}^{\rm gpu}),
    \label{eq:gpu_slice}
\end{equation}
where $\Omega_{i,a}^{\rm gpu}$ is compute capacity, $M_{i,a}^{\rm vram}$ is video memory, $H_{i,a}^{\rm bw}$ is memory bandwidth, and $\omega_{i,a}^{\rm gpu}$ is reconfiguration overhead. GPU feasibility requires
\begin{equation}
\begin{cases}
q_{u,t}^{\rm gpu,ref}\le \Omega_{i,a}^{\rm gpu},\\
q_{u,i,t}^{\rm vram}\le M_{i,a}^{\rm vram},\\
h_{u}^{\rm mem}\le H_{i,a}^{\rm bw}.
\end{cases}
\label{eq:gpu_feasible}
\end{equation}
For NPU execution, the operator and precision compatibility condition is
\begin{equation}
    \mathbbm 1_{u,i}^{\rm npu}=\mathbbm 1\{\mathcal O_u\subseteq\mathcal O_i^{\rm npu},
    \pi_u\in\mathcal Q_i^{\rm npu},q_{u,t}^{\rm sram,ref}\le M_i^{\rm sram}\}.
    \label{eq:npu_feasible}
\end{equation}
If $\mathbbm 1_{u,i}^{\rm npu}=0$, the NPU option is marked infeasible for the pair $(u,i)$.

\subsection{Cache-Energy Model}
The cache demand is decomposed as
\begin{equation}
    q_{u,i,t}^{\rm cache}=q_{u,t}^{\rm in}+q_{u,t}^{\rm mid}
    +q_{u,i,t}^{\rm model}+q_{u,t}^{\rm state}.
    \label{eq:cache_demand}
\end{equation}
The input component is $q_{u,t}^{\rm in}=B_u^{\rm ul}\Delta_t/8$ in megabytes, and the intermediate-data component is
\begin{equation}
    q_{u,t}^{\rm mid}=\sum_{p\in\mathcal C(\psi_u)}
    \kappa_p^{\rm cache}\frac{D_{u,t}^{p}}{8\times10^6}.
    \label{eq:intermediate_cache}
\end{equation}
The model-cache demand is
\begin{equation}
    q_{u,i,t}^{\rm model}=\sum_{h\in\mathcal H_u}s_h(1-\ell_{i,h,t}),
    \label{eq:model_cache}
\end{equation}
where $\mathcal H_u$ is the set of models required by user $u$, $s_h$ is the size of model $h$, and $\ell_{i,h,t}$ indicates whether it is already cached at LEO $i$. The state component is 16 MB for communication/video services and 64 MB for AI services. The storage demand used in the experiments is
\begin{equation}
\begin{aligned}
q_{u,i,t}^{\rm sto}={}&q_{u,t}^{\rm in}+q_{u,t}^{\rm out}
+0.10(q_{u,t}^{\rm in}+q_{u,t}^{\rm out})\\
&+q_{u,i,t}^{\rm model},
\end{aligned}
\label{eq:storage_demand}
\end{equation}
where $q_{u,t}^{\rm out}=B_u^{\rm dl}\Delta_t/8$ and the 10\% term represents temporary storage. The data movement delay is
\begin{equation}
    T_{u,i,t}^{\rm mov}=\frac{q_{u,t}^{\rm in}+q_{u,t}^{\rm mid}+q_{u,i,t}^{\rm model}}
    {H_i^{\rm mem}}.
    \label{eq:data_movement}
\end{equation}
where $H_i^{\rm mem}$ is the effective memory bandwidth available at LEO $i$. The energy demand is
\begin{equation}
    q_{u,i,t}^{\rm ene}=E_{u,i,t}^{\rm com}+E_{u,i,t}^{\rm cpu}
    +E_{u,i,t}^{\rm gpu}+E_{u,i,t}^{\rm npu}+E_{u,i,t}^{\rm cache}.
    \label{eq:energy_total}
\end{equation}
The executable scenario uses
\begin{equation}
\begin{aligned}
E_{u,i,t}^{\rm com}&=0.03B_u^{\rm req}\Delta_t
+0.2d_{u,i,t}^{\rm slant}/10^6,\\
E_{u,i,t}^{\rm cpu}&=10^{-11}q_{u,i,t}^{\rm cpu},\\
E_{u,i,t}^{\rm gpu}&=P_{i,g}^{\rm gpu}T_{u,i,g,t},\qquad
E_{u,i,t}^{\rm npu}=P_i^{\rm npu}T_{u,i,\rm npu,t},\\
E_{u,i,t}^{\rm cache}&=0.002q_{u,i,t}^{\rm cache}.
\end{aligned}
\label{eq:energy_terms}
\end{equation}
where slant range is in meters, execution time is in seconds, and $P_{i,g}^{\rm gpu}$ and $P_i^{\rm npu}$ are scenario-assigned power ratings in watts. The energy-conversion coefficients in~\eqref{eq:energy_terms}, accelerator power values, and energy capacities are synthetic scenario parameters; energy demands and capacities share a normalized joule-equivalent scale, without flight-hardware calibration. The assignment-induced dimensionless thermal increment is
\begin{equation}
\Delta\vartheta_{u,i,t}=\frac{q_{u,i,t}^{\rm ene}}{C_{i,\rm ene}^{\rm tot}},
\label{eq:thermal_increment}
\end{equation}
where $C_{i,\rm ene}^{\rm tot}$ is the scenario energy-capacity scale of LEO $i$. The aggregate thermal constraint is
\begin{equation}
    \vartheta_{i,t}+\sum_{u\in\mathcal U_t}x_{u,i,t}\Delta\vartheta_{u,i,t}
    \le \vartheta_i^{\max}.
    \label{eq:thermal_constraint}
\end{equation}
Here $\vartheta_{i,t}\in[0,1]$ is the normalized pre-decision thermal state and $\vartheta_i^{\max}=1$ in the experiments.
The corresponding pre-decision thermal headroom is
\begin{equation}
    H_{i,t}^{\rm th}=\vartheta_i^{\max}-\vartheta_{i,t}.
    \label{eq:thermal_headroom}
\end{equation}
With a fresh report, $\hat H_{i,t}^{\rm th}=H_{i,t}^{\rm th}$. Under report age $\zeta_{i,t}^{\rm age}>0$, the controller uses the latest available value $\hat H_{i,t-\zeta_{i,t}^{\rm age}}^{\rm th}$ as specified in~\eqref{eq:meo_view}.

\section{Problem Formulation}
\label{sec:problem}

\subsection{Target-LEO Resource Mapping}
The service-primitive-driven mapping function is
\begin{equation}
    \bm q_{u,i,t}=\mathcal F_{\Theta}
    (\bm\varrho_{u,t},i,\hat{\bm s}_{i,t},\mathcal C(\psi_u)),
    \label{eq:mapping_func}
\end{equation}
For candidate LEO $i$, the decision-time state is the LEO-specific projection of the fused global view,
\begin{equation}
    \hat{\bm s}_{i,t}=\Pi_i(\hat{\bm z}_t^{\rm G}),
    \label{eq:state_projection}
\end{equation}
where $\Pi_i(\cdot)$ extracts the capacity, link, queue, accelerator, cache, and thermal fields of candidate $i$. The state $\bm s_{i,t}^{\rm real}$ denotes post-decision execution truth used only for evaluation and calibration, and $\Theta$ denotes mapping parameters. The output vector is
\begin{equation}
\begin{aligned}
\bm q_{u,i,t}=(&q^{\rm bw},q^{\rm cpu},q^{\rm gpu},q^{\rm npu},q^{\rm vram},\\
&q^{\rm sram},q^{\rm cache},q^{\rm sto},q^{\rm ene})_{u,i,t}.
\end{aligned}
\label{eq:q_vector}
\end{equation}
For any resource $r\in\mathcal R$, the physical demand in~\eqref{eq:physical_demand_quantization} is quantized to the scheduling granularity:
\begin{equation}
    \hat q_{u,i,t}^{r}=
    \left\lceil\frac{q_{u,i,t}^{r,\rm phy}}{\delta_{i,r}^{\rm sch}}\right\rceil
    \delta_{i,r}^{\rm sch}.
    \label{eq:generic_quantization}
\end{equation}
The corresponding granularity waste is
\begin{equation}
    w_{u,i,t}^{r}=\hat q_{u,i,t}^{r}-q_{u,i,t}^{r}.
    \label{eq:generic_waste}
\end{equation}
This target-dependent quantization also reflects constellation-level design, on-demand satellite IoT coverage, and AoI/capacity-oriented metric design considerations~\cite{10495304,10680167}.

\subsection{Feasibility Model}
The end-to-end latency is
\begin{equation}
\begin{aligned}
T_{u,i,t}^{\rm e2e}=&T_{u,i,t}^{\rm com}+T_{u,i,t}^{\rm cpu}
+T_{u,i,t}^{\rm acc}\\
&+T_{u,i,t}^{\rm mov}+T_{u,i,t}^{\rm cache},
\end{aligned}
\label{eq:e2e_delay}
\end{equation}
where $T_{u,i,t}^{\rm cpu}=q_{u,i,t}^{\rm cpu}/F_{i,t}^{\rm cpu}$, $T_{u,i,t}^{\rm cache}$ is the cache lookup or model-loading delay implied by the cache state, and $T_{u,i,t}^{\rm acc}$ is GPU/NPU acceleration delay. Services without intelligent inference or accelerated computing use $T_{u,i,t}^{\rm acc}=0$. The controller-side QoS margin is
\begin{equation}
\Gamma_{u,i,t}=\min\!\left\{
\frac{R_{u,i,t}}{B_u^{\rm req}},
\frac{T_u^{\max}}{T_{u,i,t}^{\rm e2e}},
\min_{r:\hat q_{u,i,t}^{r}>0}
\frac{\hat C_{i,r,t}}{\hat q_{u,i,t}^{r}}
\right\}.
    \label{eq:qos_score_raw}
\end{equation}
The clipped QoS score is
\begin{equation}
    \bar\Gamma_{u,i,t}=\max\{0,\min\{1,\Gamma_{u,i,t}\}\}.
    \label{eq:qos_score_clip}
\end{equation}
The pair-compatibility indicator contains tests that can be evaluated independently for one user--LEO pair. Let $\chi_{u,i,t}^{\rm acc}=1$ when the required operators, numerical precision, and an accelerator slice with sufficient local memory are supported, and zero otherwise. Then
\begin{equation}
\begin{aligned}
\mathcal I^{\rm pair}_{u,i,t}
=v_{u,i,t}\mathbbm 1\{T_{u,i,t}^{\rm e2e}\le T_u^{\max}\}
\,\chi_{u,i,t}^{\rm acc}
\prod_{r\in\mathcal R}\mathbbm 1\{\hat q_{u,i,t}^{r}\le\hat C_{i,r,t}\}\\
{}\times\mathbbm 1\{\Delta\vartheta_{u,i,t}\le \hat H_{i,t}^{\rm th}\}.
\end{aligned}
\label{eq:feas_indicator}
\end{equation}
Equation~\eqref{eq:feas_indicator} removes a candidate that cannot host even the single request under the reported state. Shared capacity is subsequently imposed on the complete assignment through~\eqref{eq:problem_resource}. GPU/NPU slices and their local VRAM/SRAM are expanded into resource atoms, so conflicts among several users enter the aggregate constraints. The pre-decision queue and cache states are frozen in $T^{\rm e2e}_{u,i,t}$ and $\hat{\bm q}_{u,i,t}$, whereas cache-load and thermal increments are charged to candidates and summed over selected users.

\subsection{Objective Metrics}
The intrinsic service value is defined when the request enters the controller as
\begin{equation}
	V_{u,t}=p_u\eta_uQ_u,
	\label{eq:service_value}
\end{equation}
where $p_u$, $\eta_u$, and $Q_u$ are priority, reliability, and requested quality. This value is counted once in the primary objective. The margin $\bar\Gamma_{u,i,t}$ reports candidate headroom, while deadline and compatibility enter the hard pair screen~\eqref{eq:feas_indicator}. The pair-specific signaling term is $s_{u,i,t}=b^{\rm cmd}_{u,i,t}$, the number of bits required to identify the target and encode its quantized allocation command. Fixed telemetry reporting is represented separately by $S_t^{\rm rep}$. The primary value of pair $(u,i)$ is
\begin{equation}
	a_{u,i,t}=V_{u,t}\mathcal I^{\rm pair}_{u,i,t}.
\label{eq:pair_value}
\end{equation}
The slot objective is
\begin{equation}
	J_t(\bm x_t)=\sum_{u\in\mathcal U_t}\sum_{i\in\mathcal S}
	x_{u,i,t}a_{u,i,t}.
    \label{eq:capacity_single}
\end{equation}
Thus, the formal problem, exact predicted-state MILP, and CAP-PD-UA use the same primary service value. Waste and signaling are reported separately and provide an infinitesimal deterministic tie-break in CAP-PD-UA. Let $e_{u,i,t}^{\rm real}$ indicate successful post-decision execution. The realized service utility is $J_t^{\rm real}=\sum_{u,i}x_{u,i,t}V_{u,t}e_{u,i,t}^{\rm real}$. This distinction yields two noninterchangeable ratios:
\begin{equation}
	\varphi_t^{\rm alg}=\frac{J_t(\bm x_t^{\rm CAP})}{J_t^{\rm pred,*}},
	\qquad
	\varphi_t^{\rm e2e}=\frac{J_t^{\rm real}(\bm x_t^{\rm CAP})}{J_t^{\rm truth,*}},
    \label{eq:capacity_ratio_gap}
\end{equation}
where $J_t^{\rm pred,*}$ is the exact MILP optimum on the controller view and $J_t^{\rm truth,*}$ is an offline oracle MILP on hidden execution truth. The first ratio isolates association quality; the second also includes inference error.

Outage accounting includes unassigned requests. Let $N_t^{\rm att}=\sum_{u,i}x_{u,i,t}$ and let $N_t^{\rm srv}$ be the number that finish under execution truth. We report
\begin{equation}
	P_t^{\rm e2e}=1-\frac{N_t^{\rm srv}}{|\mathcal U_t|},\quad
	P_t^{\rm exe}=1-\frac{N_t^{\rm srv}}{N_t^{\rm att}+\varepsilon},\quad
	A_t=\frac{N_t^{\rm att}}{|\mathcal U_t|}.
    \label{eq:outage_prob}
\end{equation}
These are end-to-end outage, conditional execution outage, and assignment coverage, respectively. Reporting all three prevents a conservative method from appearing reliable merely by rejecting users.

The resource consumption and access efficiency are
\begin{equation}
\begin{aligned}
R_t^{\rm cc}&=\sum_{u,i,r}x_{u,i,t}c_r^{\rm norm}\hat q_{u,i,t}^{r},\\
\eta_t^{\rm acc}&=\frac{J_t^{\rm real}}{R_t^{\rm cc}+\varepsilon},
\end{aligned}
\label{eq:efficiency}
\end{equation}
where $c_r^{\rm norm}$ is the unit-normalization coefficient for resource $r$.

\subsection{Signaling Model}
The state-reporting overhead is
\begin{equation}
    S_t^{\rm rep}=\sum_{i\in\mathcal S}\sum_{r\in\mathcal R}
    b_r^{\rm rep}\frac{\Delta_t}{\Delta_{i,r}^{\rm rep}},
    \label{eq:report_overhead}
\end{equation}
where $b_r^{\rm rep}$ is the number of reporting bits and $\Delta_{i,r}^{\rm rep}$ is the reporting interval. The association-command overhead is
\begin{equation}
    S_t^{\rm cmd}=\sum_{u,i}x_{u,i,t}b^{\rm cmd}_{u,i,t}
    =\sum_{u,i}x_{u,i,t}s_{u,i,t}.
    \label{eq:cmd_overhead}
\end{equation}
The total signaling overhead and capacity-signaling efficiency are
\begin{equation}
    S_t=S_t^{\rm rep}+S_t^{\rm cmd},\qquad
    \eta_t^{\rm sig}=\frac{J_t^{\rm real}}{S_t+\varepsilon}.
    \label{eq:sig_eff}
\end{equation}

\subsection{Capacity-Oriented Association Problem}
The association problem is formulated as
\begin{subequations}
\label{eq:problem}
\begin{align}
\max_{\bm x_t}\quad &J_t(\bm x_t) \label{eq:problem_obj}\\
\text{s.t.}\quad
&\sum_{i\in\mathcal S}x_{u,i,t}\le 1, &&\forall u,\label{eq:problem_unique}\\
&x_{u,i,t}\le \mathcal I^{\rm pair}_{u,i,t}, &&\forall u,i,\label{eq:problem_visible}\\
&\sum_{u\in\mathcal U_t}x_{u,i,t}\hat q_{u,i,t}^{r}\le \hat C_{i,r,t},
&&\forall i,r,\label{eq:problem_resource}\\
&\sum_u x_{u,i,t}\Delta\vartheta_{u,i,t}
\le \hat H_{i,t}^{\rm th}, &&\forall i,\label{eq:problem_thermal}\\
&x_{u,i,t}\in\{0,1\}, &&\forall u,i.\label{eq:problem_binary}
\end{align}
\end{subequations}
Compatibility is embedded only in $\mathcal I^{\rm pair}_{u,i,t}$; all shared bandwidth, accelerator, memory, cache, storage, energy, and thermal limits are enforced jointly. Problem~\eqref{eq:problem} is a multidimensional generalized assignment problem with discrete association variables and target-dependent demands and is NP-hard.

Across epochs, Problem~\eqref{eq:problem} carries session state forward. Let $n_{u,t}$ indicate a new request, let $\ell_{u,t}$ be the remaining slots before the decision, and define $\delta_{u,t}^{\rm act}=\mathbbm 1\{\ell_{u,t}>0\}$. Request identities make $\delta_{u,t}^{\rm act}n_{u,t}=0$. Since $n_{u,t}$ is fixed at the epoch, $b_{u,t}=n_{u,t}\sum_i x_{u,i,t}$ records first-slot admission without a bilinear decision term. For $i\ne j$, the temporal accounting is
\begin{equation}
\begin{aligned}
\sum_i x_{u,i,t}&\le \delta_{u,t}^{\rm act}+n_{u,t},\\
\ell_{u,t+1}&=\big[\ell_{u,t}+b_{u,t}d_u-\sum_i x_{u,i,t}\big]^+,\\
x_{u,i,t-1}+x_{u,j,t}-1&\le y_{u,i,j,t}
\le\min\{x_{u,i,t-1},x_{u,j,t}\},\\
H_t&=\sum_{u,i\ne j}y_{u,i,j,t},\quad y_{u,i,j,t}\in\{0,1\}.
\end{aligned}
\label{eq:session_persistence}
\end{equation}
An admitted session is reconsidered in each slot until $\ell_{u,t}=0$ or the consecutive-unserved drop threshold is reached. Whenever it is assigned, its mapped resources are charged in~\eqref{eq:problem_resource}; $H_t$ triggers the handover delay and state-transfer demand already included in the mapped value. Slotwise visibility confines assignment to the current contact window under uncertain future visibility. The method is a receding one-epoch policy with persistence, resource holding during service, and handover accounting. Predictive slice re-embedding provides a complementary horizon-planning extension; the temporal experiment applies~\eqref{eq:session_persistence} and reports blocking, drops, and handovers caused by limited contact continuity.

\section{Proposed Scheme}
\label{sec:scheme}

\begin{figure*}[t]
	\centering
	\includegraphics[width=\textwidth]{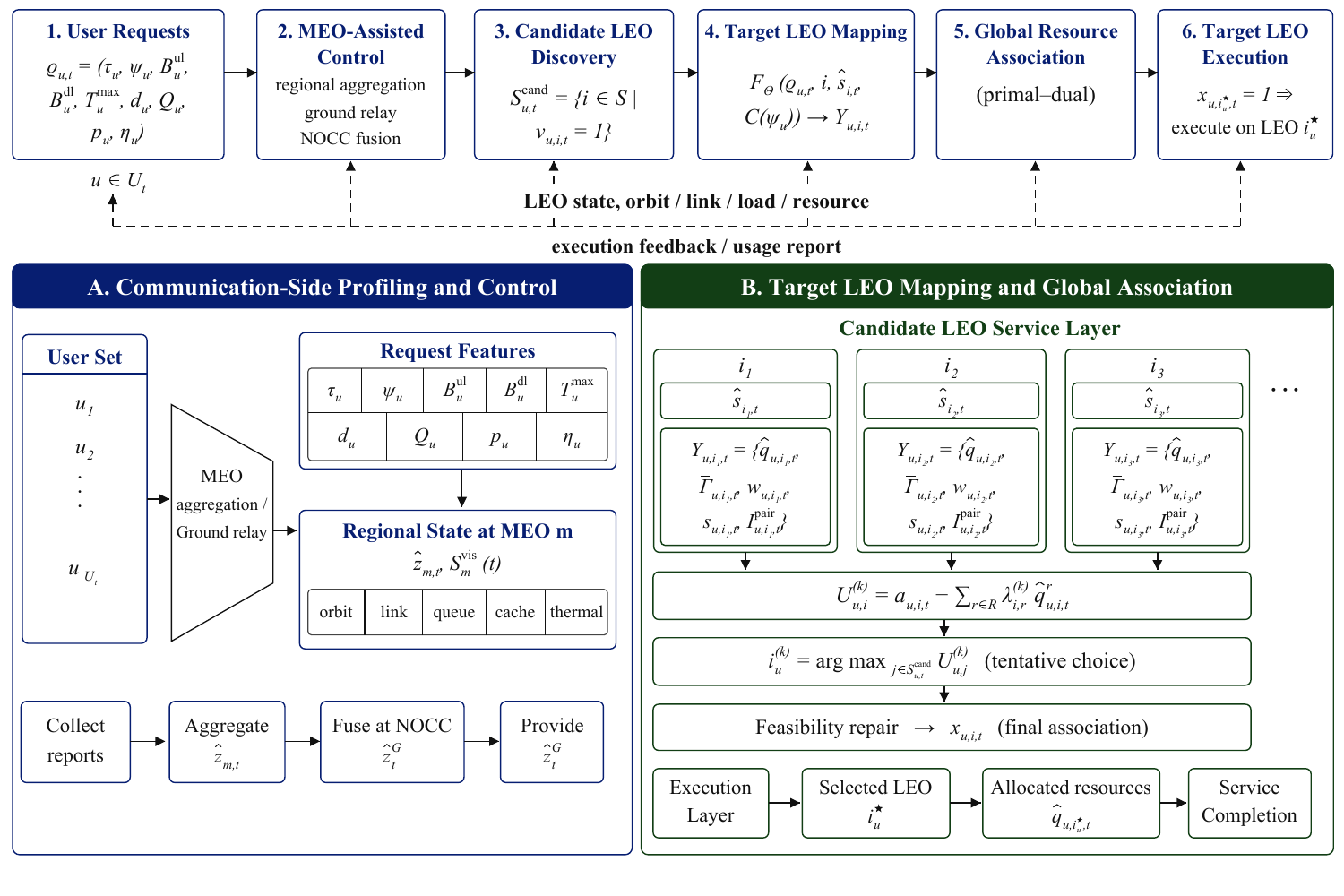}
	\caption{End-to-end workflow of the proposed MEO-assisted mapping-aware association scheme. Communication-side profiling aggregates user requests and regional LEO state, while target-dependent resource mapping evaluates candidate service resources and the NOCC selects the execution LEO. Solid and dashed arrows denote the main decision flow and execution feedback, respectively.}
	\label{fig:scheme_flow}
\end{figure*}

\subsection{Scheme Overview}

Fig.~\ref{fig:scheme_flow} follows the six stages from request arrival to service execution. Each request is represented by bandwidth, latency, reliability, and service-type fields. MEO regional aggregation, ground relay, and NOCC global synchronization combine these fields with time-stamped orbit, link, load, and resource state. Visible candidates are then discovered, and SP-CCRM and TL-QRM convert the request into a candidate-specific communication--computing resource vector. The NOCC evaluates utility, delay, energy, handover cost, intrinsic compatibility, and shared capacities before issuing the binary association command. The maximization shown in the association block is the tentative iteration-$k$ choice; Algorithm~\ref{alg:cappd} subsequently repairs aggregate violations and returns the final $x_{u,i,t}$. In the execution layer, $i_u^\star$ denotes the final selected LEO satisfying $x_{u,i_u^\star,t}=1$. The selected LEO allocates compute, memory, storage, and energy resources and returns usage and execution feedback for subsequent state aggregation and calibration. Thus, MEO satellites support regional sensing and relay, whereas the NOCC remains the global logical decision authority. Fine-grained mapped demands extend satellite MEC pricing, SAG offloading, multitask satellite-IoT MEC, and NTN implementation beyond bandwidth or aggregate AI-TOPS state~\cite{9448065,9928786,10024305,11073184}.

\subsection{Target-LEO Mapping}

The primitive profile $\Theta=\{\bm\kappa_p^r,\gamma_p,\kappa_p^{\rm cache},\delta_{i,r}^{\rm phy},\delta_{i,r}^{\rm sch},\delta_{i,r}^{\rm rep}\}$ is initialized before association. For a compute resource $r\in\mathcal R_{\rm cmp}=\{\mathrm{cpu},\mathrm{gpu},\mathrm{npu}\}$, $\bm\kappa_p^r$ contains the per-bit, per-packet, per-frame, per-token, and per-request coefficients of primitive $p$; entries unsupported by a primitive--resource pair are zero. Trace fields supply time, location, and bandwidth, while primitive graphs determine event counts, processed bytes, and output/input ratios. Per-resource work is divided by reference per-slot throughput and expressed in scheduler quanta. Candidate-specific cache misses, model loading, data movement, reconfiguration, and energy enter the mapped demand through~\eqref{eq:cache_demand}--\eqref{eq:energy_terms} and the accelerator-selection rules. This operation-count/throughput normalization follows public inference-benchmark methodology~\cite{reddi2020mlperf}. Profiles are fixed before traffic seeds and shared by all methods.

The three $\delta$ terms denote hardware atoms, scheduler quanta, and telemetry encoding. In the absence of flight-computer telemetry, the catalog is a dimensionless hardware-class envelope: a reference coefficient converts its associated request feature into scheduler work. The $0.5$--$4$ sweep scales every nonzero entry of $\bm\kappa_p^r$ while preserving $\gamma_p$, $\kappa_p^{\rm cache}$, and primitive-chain composition. Hidden execution perturbation is varied up to 30\%, and static, coarse, oracle, and calibrated mappings are compared. These experiments characterize robustness to scale and mismatch within the stated hardware-class envelope.

TL-QRM is the resource-mapping interface between the service model and the association algorithm. For each candidate pair $(u,i)$, it receives the request $\bm\varrho_{u,t}$, the delayed target estimate $\hat{\bm s}_{i,t}$, and $\Theta$, and outputs
\begin{equation}
	\mathcal Y_{u,i,t}
	=
	\{\hat{\bm q}_{u,i,t},\bar\Gamma_{u,i,t},\bm w_{u,i,t},
	s_{u,i,t},\mathcal I^{\rm pair}_{u,i,t}\},
	\label{eq:tlqrm_output}
\end{equation}
where the last element represents the single-request pair screen; aggregate feasibility is enforced by~\eqref{eq:problem_resource}. The algorithm contains four logical layers: service-primitive expansion, continuous demand estimation, target-LEO-aware quantization, and compatibility/QoS evaluation.

First, TL-QRM expands the service type $\psi_u$ into a primitive chain $\mathcal C(\psi_u)$ and computes the data-volume evolution according to~\eqref{eq:data_evolution}. Define the feature vector processed by primitive $p$ as
\begin{equation}
	\bm d_{u,t}^{p}=
	(D_{u,t}^{p},N_{u,t}^{\rm pkt},N_{u,t}^{\rm frm},N_{u,t}^{\rm tok},1)^{\mathsf T}.
	\label{eq:primitive_feature}
\end{equation}
The service-intrinsic reference workload of compute resource $r\in\mathcal R_{\rm cmp}$ is
\begin{equation}
	q_{u,t}^{r,\rm ref}
	=\sum_{p\in\mathcal C(\psi_u)}
	\langle\bm\kappa_{p}^{r},\bm d_{u,t}^{p}\rangle.
	\label{eq:tlqrm_continuous_demand}
\end{equation}
TL-QRM then forms each candidate-dependent compute demand as
\begin{equation}
    q_{u,i,t}^{r}=z_{u,i,t}^{r}q_{u,t}^{r,\rm ref}
    +\Delta q_{u,i,t}^{r},
    \label{eq:target_correction}
\end{equation}
for $r\in\mathcal R_{\rm cmp}$, where $z_{u,i,t}^{\rm cpu}=1$ and $z_{u,i,t}^{r}$ indicates the selected implementation for $r\in\{\mathrm{gpu},\mathrm{npu}\}$. The correction $\Delta q_{u,i,t}^{r}$ contains candidate-specific compute overhead and is zero on the unselected accelerator. Communication, cache, storage, and energy demands follow~\eqref{eq:comm_alloc}, \eqref{eq:cache_demand}, \eqref{eq:storage_demand}, and~\eqref{eq:energy_total}, respectively. Together with the link, queue, granularity, accelerator, and thermal rules, these terms produce the target dependence evaluated in the experiments.

Second, continuous demands are converted into scheduling blocks. The communication and CPU block mappings are
\begin{equation}
	\begin{aligned}
		n_{u,i,t}^{\rm bw}&=\left\lceil
		\frac{q_{u,i,t}^{\rm bw,phy}}{\delta_{i,\mathrm{bw}}^{\rm sch}}
		\right\rceil,\\
		n_{u,i,t}^{\rm cpu}&=\left\lceil
		\frac{q_{u,i,t}^{\rm cpu,phy}}{\delta_{i,\mathrm{cpu}}^{\rm sch}}
		\right\rceil.
	\end{aligned}
	\label{eq:block_mapping_scheme}
\end{equation}
For all resources, TL-QRM applies the physical and scheduling rules in~\eqref{eq:physical_demand_quantization} and~\eqref{eq:generic_quantization}. The resulting total granularity waste is
\begin{equation}
	w_{u,i,t}^{r}=\hat q_{u,i,t}^{r}-q_{u,i,t}^{r}.
	\label{eq:tlqrm_quant_waste}
\end{equation}
Thus, TL-QRM explicitly accounts for resource granularity and allocation waste before the association decision.

Third, accelerator-related primitives trigger GPU/NPU feasibility checks. For accelerator option $g$, $T_{u,i,g,t}$, $E_{u,i,g,t}$, and $W_{u,i,g,t}$ denote predicted delay, energy, and granularity waste; $\lambda_T$, $\lambda_E$, and $\lambda_W$ are their normalized nonnegative weights. If GPU is triggered, TL-QRM constructs the feasible GPU-slice set $\mathcal G_{i,t}^{\rm fea}(u)$ according to compute capacity, video-memory capacity, and memory-bandwidth constraints, and then selects
\begin{equation}
	g_{u,i,t}^{\rm gpu}
	=
	\mathop{\arg\min}_{g\in\mathcal G_{i,t}^{\rm fea}(u)}
	\big(\lambda_TT_{u,i,g,t}+\lambda_EE_{u,i,g,t}+\lambda_WW_{u,i,g,t}\big).
	\label{eq:gpu_selection}
\end{equation}
For NPU execution, TL-QRM checks operator support and precision compatibility. If both GPU and NPU are feasible, the selected accelerator is
\begin{equation}
	g_{u,i,t}^{\star}
	=
	\mathop{\arg\min}_{g\in\{\mathrm{gpu},\mathrm{npu}\}}
	\big(\lambda_TT_{u,i,g,t}+\lambda_EE_{u,i,g,t}+\lambda_WW_{u,i,g,t}\big).
	\label{eq:acc_selection}
\end{equation}
Finally, TL-QRM evaluates end-to-end delay, accelerator compatibility, and the necessary single-request fit to obtain $\bar\Gamma_{u,i,t}$ and $\mathcal I^{\rm pair}_{u,i,t}$. Aggregate cache, energy, accelerator-unit, and thermal limits are evaluated after candidate pairs are combined by the association problem.

Algorithm~\ref{alg:tlqrm} summarizes primitive-level demand estimation, target-specific quantization, and feasibility checking over communication, compute, memory, storage, energy, and thermal constraints.

\begin{algorithm}[!t]
	\caption{TL-QRM: Target-LEO Quantized Resource Mapping}
	\label{alg:tlqrm}
	\begin{algorithmic}[1]
		\REQUIRE User request $\bm\varrho_{u,t}$, candidate LEO $i$, target-state estimate $\hat{\bm s}_{i,t}$, primitive profiles $\Theta$.
		\ENSURE Mapping output $\mathcal Y_{u,i,t}=\{\hat{\bm q}_{u,i,t},\bar\Gamma_{u,i,t},\bm w_{u,i,t},s_{u,i,t},\mathcal I^{\rm pair}_{u,i,t}\}$.
		\STATE \textbf{Primitive expansion:} Generate $\mathcal C(\psi_u)$ and compute $D_{u,t}^{p}$ for all $p\in\mathcal C(\psi_u)$ using~\eqref{eq:data_evolution}.
		\STATE \textbf{Continuous demand estimation:} Compute the reference compute demands by~\eqref{eq:tlqrm_continuous_demand}; obtain each target-dependent $q_{u,i,t}^{r}$ from~\eqref{eq:target_correction} and the explicit communication, cache, storage, energy, and thermal rules.
	\STATE \textbf{Communication mapping:} Compute $T_{u,i,t}^{\rm com}$ by~\eqref{eq:comm_delay} and obtain $n_{u,i,t}^{\rm bw}$ by~\eqref{eq:block_mapping_scheme}.
	\STATE \textbf{CPU mapping:} Compute $q_{u,i,t}^{\rm cpu}$ and obtain $n_{u,i,t}^{\rm cpu}$ by~\eqref{eq:block_mapping_scheme}.
		\IF{GPU-related primitives are triggered}
		\STATE Construct $\mathcal G_i^{\rm fea}(u)$ by checking compute, video-memory, and memory-bandwidth constraints.
		\STATE Select $g_{u,i,t}^{\rm gpu}$ according to~\eqref{eq:gpu_selection}.
		\ENDIF
		\IF{NPU-related primitives are triggered}
		\STATE Check operator support and precision compatibility.
		\ENDIF
		\IF{both GPU and NPU are feasible}
		\STATE Select $g_{u,i,t}^{\star}$ according to~\eqref{eq:acc_selection}.
		\ENDIF
		\STATE \textbf{Cache/storage/energy mapping:} Compute cache miss, model-loading demand, data-movement delay, storage demand, energy demand, and thermal risk.
		\STATE \textbf{Quantization:} For each $r\in\mathcal R$, compute $q_{u,i,t}^{r,\rm phy}$, $\hat q_{u,i,t}^{r}$, and $w_{u,i,t}^{r}$ using~\eqref{eq:physical_demand_quantization}, \eqref{eq:generic_quantization}, and~\eqref{eq:tlqrm_quant_waste}.
	\STATE \textbf{QoS and compatibility:} Compute $\bar\Gamma_{u,i,t}$ and $\mathcal I^{\rm pair}_{u,i,t}$ from link/delay margins, accelerator compatibility, single-request fit, and thermal headroom.
		\STATE \textbf{Signaling accounting:} Set $s_{u,i,t}=b^{\rm cmd}_{u,i,t}$; account fixed state reporting separately through~\eqref{eq:report_overhead}.
		\STATE \textbf{return} $\mathcal Y_{u,i,t}$.
	\end{algorithmic}
\end{algorithm}

If each user has on average $K$ candidate LEOs, the average number of GPU slices is $G$, the number of NPU tiles is $N$, and the primitive-chain length is $L$, the mapping complexity is
\begin{equation}
	\mathcal O\big(|\mathcal U_t|K(L+G+N)\big).
	\label{eq:tlqrm_complexity}
\end{equation}

\subsection{Primal--Dual Association}

Given the TL-QRM outputs $\{\mathcal Y_{u,i,t}\}$, CAP-PD-UA converts multidimensional scarcity into prices using $a_{u,i,t}$, $\hat{\bm q}_{u,i,t}$, $\bm w_{u,i,t}$, $\mathcal I^{\rm pair}_{u,i,t}$, and $\hat C_{i,r,t}$. The margin $\bar\Gamma_{u,i,t}$ remains a reported candidate-headroom diagnostic.

Let $\lambda_{i,r}^{(k)}$ be the price of resource $r$ on LEO $i$ at iteration $k$. The utility of associating user $u$ with LEO $i$ is
\begin{equation}
	\begin{aligned}
		U_{u,i}^{(k)}=a_{u,i,t}
		-\sum_{r\in\mathcal R}\lambda_{i,r}^{(k)}\hat q_{u,i,t}^{r}.
	\end{aligned}
	\label{eq:utility}
\end{equation}
Here $a_{u,i,t}$ is exactly the coefficient in the formal objective~\eqref{eq:pair_value}; the price term is the Lagrangian relaxation of the aggregate resource-capacity constraints in~\eqref{eq:problem_resource}. The implementation uses $\varepsilon_{\rm tie}(\sum_r w_{u,i,t}^r/(\hat C_{i,r,t}+\varepsilon)+s_{u,i,t}/10^4)$, with $0<\varepsilon_{\rm tie}\ll1$, as a deterministic secondary tie-break and charges the handover term in a persistent session. These secondary terms leave the exact-MILP audit and all reported primary objectives on the service-value definition in~\eqref{eq:service_value}. For each user, the NOCC selects the candidate LEO with the maximum positive utility:
\begin{equation}
	i_u^{(k)}
	=
	\mathop{\arg\max}_{i\in\mathcal S_{u,t}^{\rm cand}}
	U_{u,i}^{(k)}.
	\label{eq:user_select}
\end{equation}
If the maximum utility is non-positive, user $u$ remains unassigned in the current iteration.

After constructing the tentative association, the NOCC updates the price indexed by each LEO--resource pair according to overload:
\begin{equation}
	\lambda_{i,r}^{(k+1)}
	=
	\Big[
	\lambda_{i,r}^{(k)}
	+
	\rho_k
	\Big(
	\sum_u x_{u,i,t}^{(k)}\hat q_{u,i,t}^{r}
	-
		\hat C_{i,r,t}
	\Big)
	\Big]^+.
	\label{eq:price_update}
\end{equation}
The default finite-horizon schedule is $\rho_k=\rho_0/\sqrt{k+1}$ over at most 50 iterations. A positive overload increases the corresponding resource price. The discrete algorithm terminates when the projected assignment becomes stable and satisfies the resource and thermal constraints or when $K_{\max}$ is reached, after which feasibility repair is applied. Its finite-sample quality is evaluated by exact-MILP ratios and the convergence sweep; no asymptotic claim is required for the repaired integer iterate.

The primal--dual iterations may still leave a small number of infeasible assignments because resources are discrete and multiple constraints are coupled. Therefore, CAP-PD-UA performs feasibility repair after price convergence. For each overloaded satellite-resource pair, the unit-resource benefit of user $u$ on LEO $i$ is
\begin{equation}
	\beta_{u,i,r}
	=
	\frac{\max\{a_{u,i,t},0\}}
	{\hat q_{u,i,t}^{r}+\varepsilon}.
	\label{eq:unit_benefit}
\end{equation}
Let $k_f$ be the last price iteration before repair and define $U_{u,j}^{\rm fin}=U_{u,j}^{(k_f)}$. For the current repair matrix $\tilde{\bm x}_t$, the residual reported resource capacity and thermal headroom available to user $u$ on candidate $j$ are $C_{j,r,t}^{\rm rem}=\hat C_{j,r,t}-\sum_{v\ne u}\tilde x_{v,j,t}\hat q_{v,j,t}^{r}$ and $H_{j,t}^{\rm th,rem}=\hat H_{j,t}^{\rm th}-\sum_{v\ne u}\tilde x_{v,j,t}\Delta\vartheta_{v,j,t}$, respectively. Users with lower $\beta_{u,i,r}$ are removed or rematched first. A removed user can be rematched to another LEO $j$ only when
\begin{equation}
\begin{aligned}
	U_{u,j}^{\rm fin}&>0,\qquad \mathcal I_{u,j,t}^{\rm pair}=1,\\
	\hat q_{u,j,t}^{r}&\le C_{j,r,t}^{\rm rem},\quad \forall r\in\mathcal R,\\
	\Delta\vartheta_{u,j,t}&\le H_{j,t}^{\rm th,rem},
\end{aligned}
	\label{eq:rematch_condition}
\end{equation}
For a thermal violation, users are ordered by the analogous score $\beta_{u,i}^{\rm th}=\max\{a_{u,i,t},0\}/(\Delta\vartheta_{u,i,t}+\varepsilon)$. Each removal strictly decreases the number of tentative assignments on an overloaded LEO. Rematching is accepted only after the pair-compatibility, residual-resource, and residual-thermal tests pass, and a user is visited at most once per candidate. Hence repair terminates after at most $|\mathcal U_t|K$ candidate visits and returns a feasible assignment; the all-unassigned assignment is the finite fallback. Exact-MILP comparisons quantify the optimality gap of this feasible finite procedure.

Algorithm~\ref{alg:cappd} combines utility-driven selection, per-LEO resource pricing, and feasibility repair to return an association satisfying all hard constraints.

\begin{algorithm}[!t]
	\caption{CAP-PD-UA: Capacity-Aware Primal--Dual User Association}
	\label{alg:cappd}
	\begin{algorithmic}[1]
	\REQUIRE Candidate mappings $\{\mathcal Y_{u,i,t}\}$, visibility $v_{u,i,t}$, reported capacities $\hat C_{i,r,t}$, reported thermal headroom $\hat H_{i,t}^{\rm th}$, maximum iteration $K_{\max}$.
		\ENSURE Feasible association matrix $\bm x_t$.
		\STATE Initialize $\lambda_{i,r}^{(0)}=0$ and $x_{u,i,t}^{(0)}=0$ for all $u,i,r$.
		\FOR{$k=0,1,\ldots,K_{\max}-1$}
		\STATE \textbf{NOCC per-user selection:} Compute $U_{u,i}^{(k)}$ by~\eqref{eq:utility} for all visible and pair-compatible candidate pairs.
		\STATE For each user, select $i_u^{(k)}$ by~\eqref{eq:user_select} if the maximum utility is positive; otherwise leave it unadmitted.
		\STATE Construct the tentative association matrix $\bm x_t^{(k)}$.
		\STATE \textbf{NOCC per-LEO pricing:} Update $\lambda_{i,r}^{(k+1)}$ by~\eqref{eq:price_update} for all $i\in\mathcal S$ and $r\in\mathcal R$.
		\IF{the tentative association is unchanged and all resource and thermal constraints are satisfied}
		\STATE break.
		\ENDIF
		\ENDFOR
		\STATE \textbf{Feasibility repair:} For every LEO violating a resource or thermal constraint, compute the corresponding unit-benefit score.
		\STATE Remove or rematch low-benefit users according to~\eqref{eq:rematch_condition} until all hard constraints are satisfied.
		\STATE \textbf{return} $\bm x_t$.
	\end{algorithmic}
\end{algorithm}

With $K_{\max}$ iterations, $K$ average candidate LEOs, and $R=|\mathcal R|$ resource types, the complexity is
\begin{equation}
	\mathcal O\big(K_{\max}|\mathcal U_t|KR+|\mathcal U_t|K\log K\big),
	\label{eq:cappd_complexity}
\end{equation}
where the first term covers utility evaluation and price updates and the second covers sorting and feasibility repair. CAP-PD-UA avoids exponential search and supports real-time association at the NOCC.

\subsection{Online Mapping Calibration}

The mapping parameters may be biased by hardware state, model implementation, cache hits, temperature, and background load. After execution, the selected LEO reports $\bm q_{u,i,t}^{\rm real}$ to the NOCC through the MEO--ground telemetry path. The mapping error is
\begin{equation}
	\bm e_{u,i,t}
	=
	\bm q_{u,i,t}^{\rm real}
	-
	\hat{\bm q}_{u,i,t}.
	\label{eq:mapping_error}
\end{equation}
For primitive $p$ and resource $r$, let $\mu_{p,t}^{r}$ be the mean clipped ratio of realized to predicted continuous demand over the successfully served slot-$t$ requests whose chains contain $p$. The multiplicative EMA used by the executable calibration is
\begin{equation}
	\bm\kappa_{p}^{r}(t+1)=
	\bm\kappa_{p}^{r}(t)
	\big[(1-\eta_{\rm cal})+\eta_{\rm cal}\mu_{p,t}^{r}\big],
	\label{eq:ema_update}
\end{equation}
where $\mu_{p,t}^{r}$ is clipped to $[0.5,2]$ and $\eta_{\rm cal}\in(0,0.5]$ is the calibration learning rate. The multiplication applies componentwise to the nonzero entries of $\bm\kappa_p^r$; $\kappa_p^{\rm cache}$ is updated by the same rule for $r=\mathrm{cache}$. Feedback from slot $t$ is applied only from slot $t+1$, preserving causality. The correction reduces persistent mapping bias and the probability of misassociation caused by underestimated demand.

\section{Experimental Results}
\label{sec:results}

\subsection{Simulation Setup}

We use anonymized real business bandwidth-request records provided by China Satellite Communications Co., Ltd. and collected from March through April 2025, together with a 1584-satellite, 550-km, 53-degree LEO shell. The operational trace supplies time, location, and requested bandwidth. Fixed, reproducible scenario rules supply service labels, duration, and primitive operation counts because these fields and flight-hardware telemetry are absent from the trace. The results characterize comparative algorithm behavior under trace-driven offered traffic and the stated synthetic execution profiles.

\begin{figure}[!t]
	\centering
	\includegraphics[width=\columnwidth]{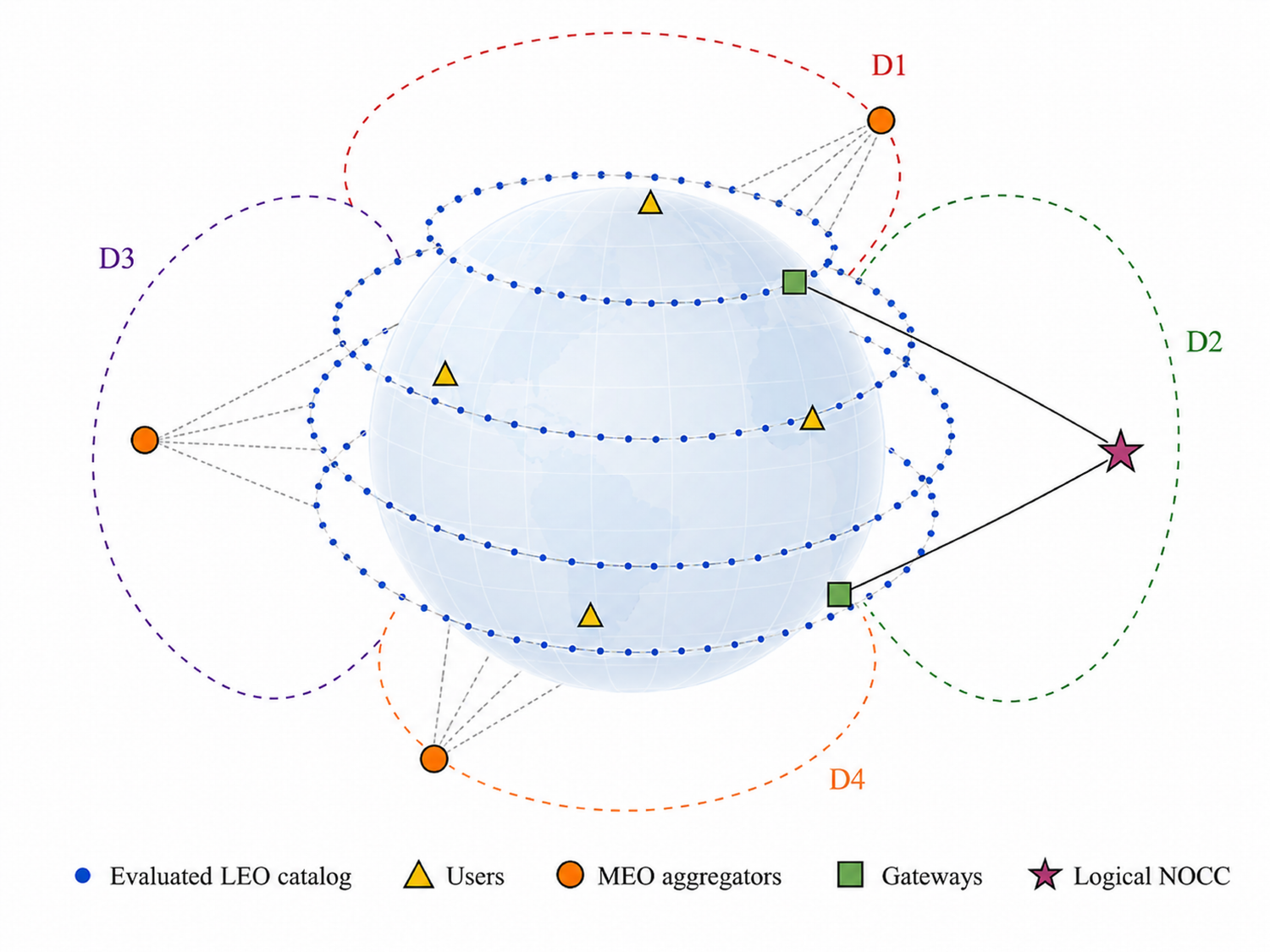}
	\caption{Representative three-dimensional view of the propagated LEO catalog and logical MEO-assisted control overlay. D1--D4 illustrate overlapping reporting domains associated with $\mathcal S_m^{\rm vis}(t)$.}
	\label{fig:architecture3d}
\end{figure}

Fig.~\ref{fig:architecture3d} visualizes the evaluated LEO geometry and the logical MEO-assisted control overlay.

Blue points form the propagated LEO catalog; yellow triangles, orange circles, green squares, and the star identify users, MEO aggregators, gateways, and the logical NOCC. The dashed D1--D4 regions illustrate overlapping reporting domains associated with $\mathcal S_m^{\rm vis}(t)$, whose membership changes with visibility. Gray telemetry links feed regional summaries to the gateways and then the NOCC for global fusion and association. The propagated LEO shell and user visibility constitute the physical orbit model. The MEO and gateway positions, MEO--LEO control links, and D1--D4 boundaries are illustrative logical-overlay elements; separate MEO orbit propagation and physical control-link budgets are reserved for deployment-level evaluation. The delay sweep evaluates control-path staleness.

The trace-driven main experiment uses seeds 1, 21, 42, and 77, with 50 stratified slots per seed and at most 120 users per slot. Each seed contributes its 50-slot mean; overall means and two-sided 95\% Student-$t$ confidence intervals use these four values with three degrees of freedom. Controlled-load and runtime sweeps vary the concurrent population from 20 to 200 users. Other sweeps use seed 7; traffic, geometry, and resources are common across methods. Hidden perturbations are revealed after association. Predicted-state and truth-state HiGHS MILPs provide the association reference and end-to-end oracle, respectively.

Table~\ref{tab:sim_params} summarizes the main simulation configuration. Table~\ref{tab:primitive_profiles} reports the primitive coefficients; $F$ and $T$ denote a frame and a token, respectively, and the last two entries are the data-volume ratio $\gamma_p$ and intermediate-cache coefficient $\kappa_p^{\rm cache}$. Omitted CPU/accelerator/cache coefficients are zero, while a data-preserving primitive has $\gamma_p=1$. Storage follows~\eqref{eq:storage_demand} and appears in the resource configuration; the primitive catalog lists the eight implemented service primitives. Table~\ref{tab:service_settings} lists the service deadline, duration, priority, reliability, and execution profile. Its second column gives $T_u^{\max}/d_u$ in seconds, and the tuple column follows the order $p_u,\eta_u$. Unless stated otherwise, $Q_u=1$; emergency requests use $p_u=3$ and $\eta_u=0.999$.

\begin{table}[!t]
	\centering
	\caption{Main Simulation Parameters}
	\label{tab:sim_params}
	\footnotesize
	\setlength{\tabcolsep}{4pt}
	\renewcommand{\arraystretch}{1.08}
	\begin{tabular}{p{0.43\columnwidth} p{0.48\columnwidth}}
		\hline
		\textbf{Parameter} & \textbf{Setting} \\
		\hline
		Traffic / main seeds & China Satcom requests, Mar.--Apr. 2025 / 1, 21, 42, 77 \\
		Slots / main users & 50 per seed / at most 120 per slot \\
		Controlled-load/runtime users & 20--200 concurrent users \\
		Service mixture & Non-emergency communication/video/AI: 40/30/30\%; emergency arrivals: 3\% \\
		Profile reference / scale & Table~\ref{tab:primitive_profiles} / $0.5$--$4$ \\
		Execution perturbation & $0$--$30\%$ of mapped demand \\
		Management overhead scale & $0$--$2$ \\
		LEO shell & 1584 satellites \\
		Orbital planes & 72 planes, 22 satellites/plane \\
		Altitude / inclination & 550 km / 53 deg \\
		Minimum elevation & 25 deg \\
		Candidate LEOs / slot & At most 6 per user / 1 s \\
		Resource dimensions & Bandwidth, CPU, GPU, NPU, VRAM, SRAM, cache, storage, energy \\
		Resource granularity & BW blocks; CPU cycles/cores; GPU slices/VRAM; NPU tiles/SRAM; cache/storage blocks; normalized joule-equivalent energy units \\
		CAP step / iterations & $\rho_k=0.2/\sqrt{k+1}$ / at most 50 \\
		LinUCB parameters & $\alpha=0.35$, ridge $=1$ \\
		RF-LinUCB parameters & $\alpha=0.30$, 48 random features \\
		MILP reference & HiGHS, 120-s limit, scaled rows \\
		Temporal run & 600 slots, 3 arrivals/slot \\
		\hline
	\end{tabular}
\end{table}

\begin{table}[!t]
	\centering
	\caption{Primitive Profiles}
	\label{tab:primitive_profiles}
	\scriptsize
	\setlength{\tabcolsep}{2.5pt}
	\renewcommand{\arraystretch}{1.02}
	\begin{tabular}{p{0.19\columnwidth}p{0.38\columnwidth}p{0.35\columnwidth}}
		\hline
		\textbf{Primitive} & \textbf{CPU coefficients} & \textbf{Accelerator; $\gamma_p$; $\kappa_p^{\rm cache}$}\\
		\hline
		Access & req.: $2.0{\times}10^5$ & --; 1; 0\\
		Transmission & bit: 2.0; pkt: 150 & --; 1; 0\\
		Security & bit: 8.0; pkt: 250 & --; 1; 0\\
		Protocol & bit: 4.0; pkt: 400 & --; 1; 0\\
		Preprocess & bit: 15; $F$: $2.5{\times}10^6$ & --; 0.65; 0.15\\
		Inference & bit: 1.0 & GPU $F$: $14{\times}10^9$, $T$: $3{\times}10^9$; NPU $F$: $6{\times}10^9$, $T$: $1{\times}10^9$; 0.08; 0.10\\
		Cache & req.: $1.0{\times}10^5$ & --; 1; 0.05\\
		Output & bit: 1.5; pkt: 80 & --; 1; 0\\
		\hline
	\end{tabular}
\end{table}

\begin{table}[!t]
	\centering
	\caption{Service Profiles}
	\label{tab:service_settings}
	\scriptsize
	\setlength{\tabcolsep}{2.2pt}
	\renewcommand{\arraystretch}{1.02}
	\begin{tabular}{p{0.18\columnwidth}p{0.21\columnwidth}p{0.24\columnwidth}p{0.29\columnwidth}}
		\hline
		\textbf{Service} & \textbf{$T_u^{\max}/d_u$ s} & \textbf{$p_u,\eta_u$} & \textbf{Profile}\\
		\hline
		Voice & 0.2/30--180 & 1.0,.990 & Comm. chain\\
		Web & 1.0/3--30 & 1.0,.990 & Comm. chain\\
		Video & 0.8/60--300 & 1.1,.990 & Comm.+cache\\
		Bulk & 5.0/20--180 & 1.1,.990 & Comm.+cache\\
		Video AI & 1.5/5--60 & 1.5,.995 & 900 MB; AI chain\\
		ASR & 1.0/2--20 & 1.5,.995 & 450 MB; AI chain\\
		LLM & 2.5/2--30 & 1.5,.995 & 5500 MB FP16\\
		Small LLM & 1.5/1--15 & 1.5,.995 & 1200 MB; AI chain\\
		Emergency AI & 0.7/10--120 & 3.0,.999 & 650 MB; full AI\\
		\hline
	\end{tabular}
\end{table}

The baselines are Max-SINR, communication-load association, linear and 48-feature random-Fourier LinUCB~\cite{li2010contextual,rahimi2007random}, and multidimensional Price-Iter. Each visible user--LEO pair is a bandit arm; the context contains the request fields, user/service one-hot codes, link state, queue, $R$ residual-resource ratios, accelerator compatibility, cache, and thermal headroom. Both bandits start without pretraining and learn from executed-service completion. All methods use identical candidates, traffic, capacity checks, and realized metrics. Oracle-Map, No-Quant, Coarse-Map, Static-LUT, and Target-Agnostic isolate the mapping layer. The exact MILP remains an offline audit.

\subsection{Main Comparison}

\begin{figure*}[!t]
	\centering
	\includegraphics[width=0.98\textwidth]{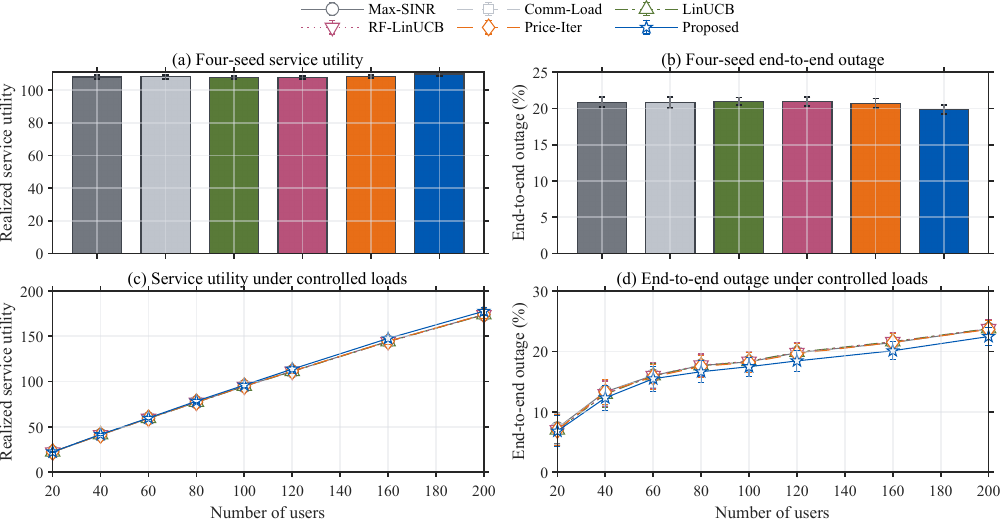}
	\caption{Overall and controlled-load comparison. Error bars show nominal two-sided 95\% Student-$t$ confidence intervals. In (a)--(b), bars average four seed-level 50-slot means, with intervals $\bar y\pm3.182s_y/\sqrt{4}$, where $s_y$ is their sample standard deviation. In (c)--(d), means and intervals use the 50 slots at each load.}
	\label{fig:main_four_seed}
\end{figure*}

Fig.~\ref{fig:main_four_seed} combines the four-seed comparison with the independent controlled-load sweep. Relative to Price-Iter, CAP-PD-UA improves utility by 1.48\%; the seed-paired utility difference is $1.597\pm0.318$, and the outage reduction is $0.901\pm0.182$ percentage points, both reported as mean $\pm$ sample standard deviation over four seeds. Both differences favor CAP-PD-UA for every seed. Light-load slots leave most methods unconstrained; separation grows with load as accelerator, memory, cache, and energy scarcity limits execution.

\begin{figure*}[!t]
	\centering
	\includegraphics[width=0.98\textwidth]{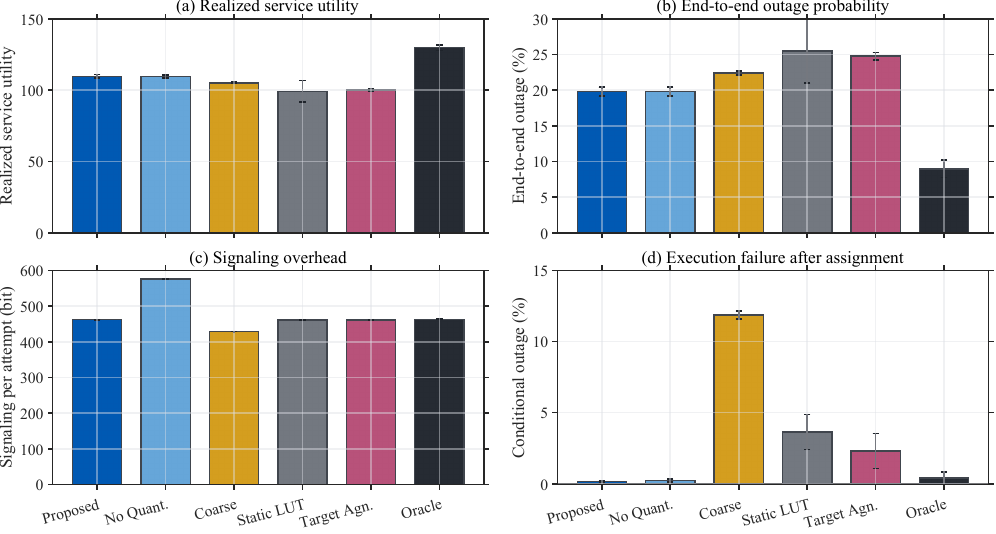}
	\caption{Mapping, oracle, static, coarse, target-agnostic, and quantization ablations under common traffic and association logic.}
	\label{fig:mapping_ablation}
\end{figure*}

\begin{figure*}[!t]
	\centering
	\includegraphics[width=0.98\textwidth]{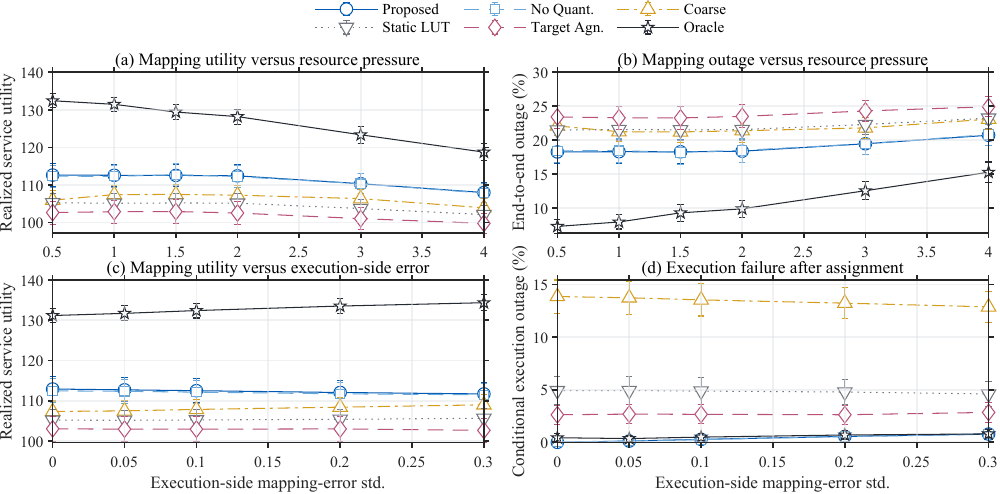}
	\caption{Mapping-variant robustness under common primitive-profile scales and hidden mapping errors.}
	\label{fig:sensitivity}
\end{figure*}

\begin{figure*}[!t]
	\centering
	\includegraphics[width=0.98\textwidth]{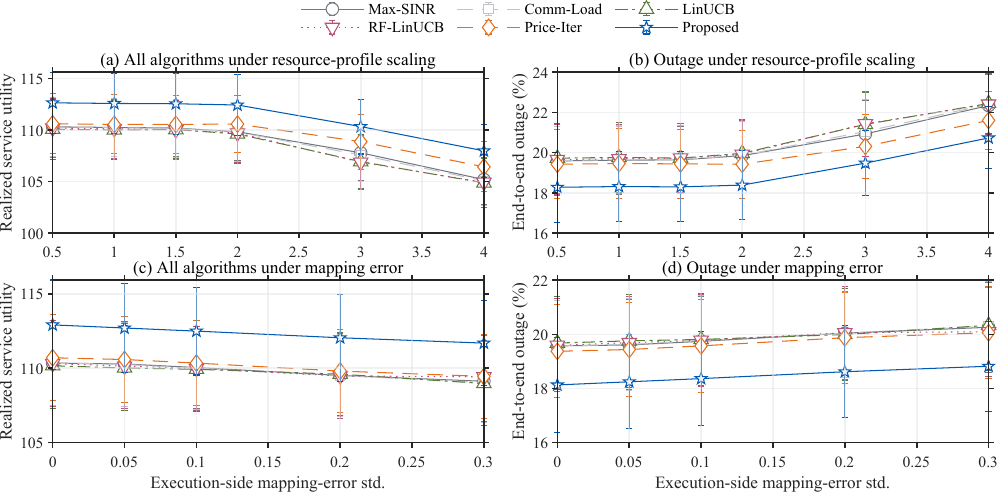}
	\caption{Association-method robustness under common primitive-profile scales and hidden mapping errors.}
	\label{fig:convergence}
\end{figure*}

\begin{figure*}[!t]
	\centering
	\includegraphics[width=0.98\textwidth]{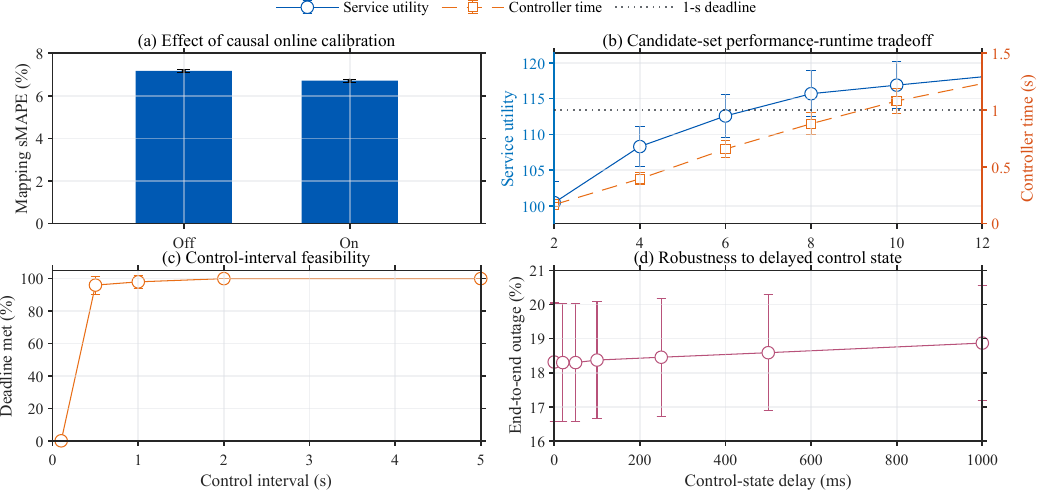}
	\caption{Candidate-set, slot-duration, control-delay, and online-calibration sensitivity.}
	\label{fig:k_slot_delay}
\end{figure*}

\begin{figure*}[!t]
	\centering
	\includegraphics[width=0.98\textwidth]{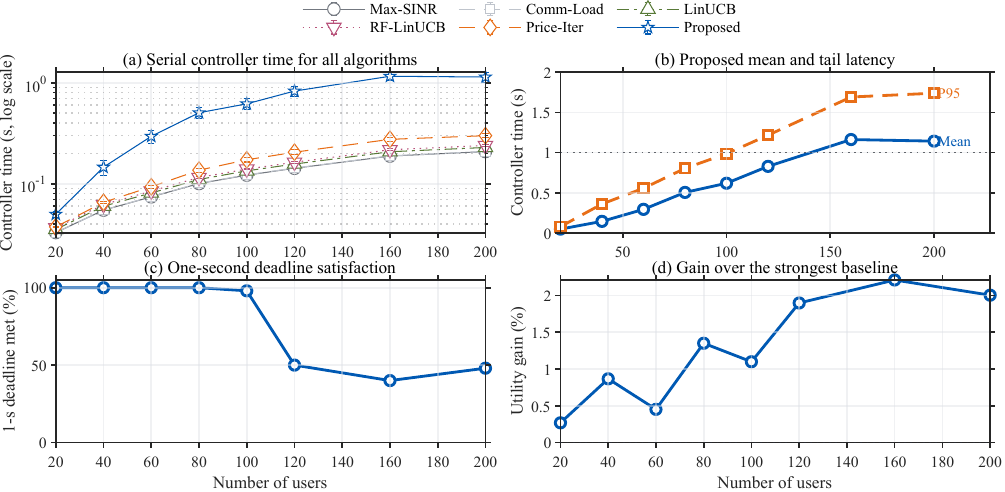}
	\caption{Serial controller-time scaling, one-second deadline satisfaction, and common-load method comparison.}
	\label{fig:runtime}
\end{figure*}

\begin{figure*}[!t]
	\centering
	\includegraphics[width=0.98\textwidth]{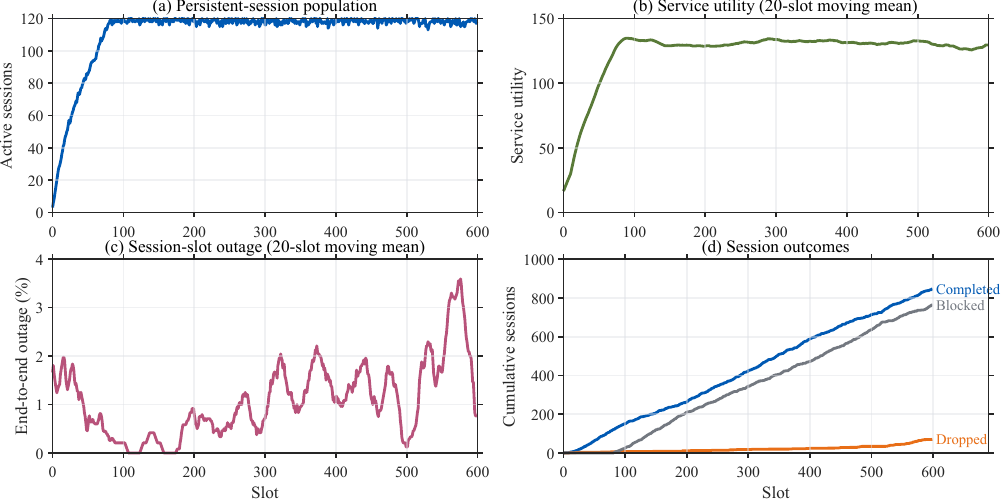}
	\caption{Persistent-session population, admission and completion outcomes, handovers, and session-slot reliability.}
	\label{fig:temporal}
\end{figure*}

\begin{figure*}[!t]
	\centering
	\includegraphics[width=0.98\textwidth]{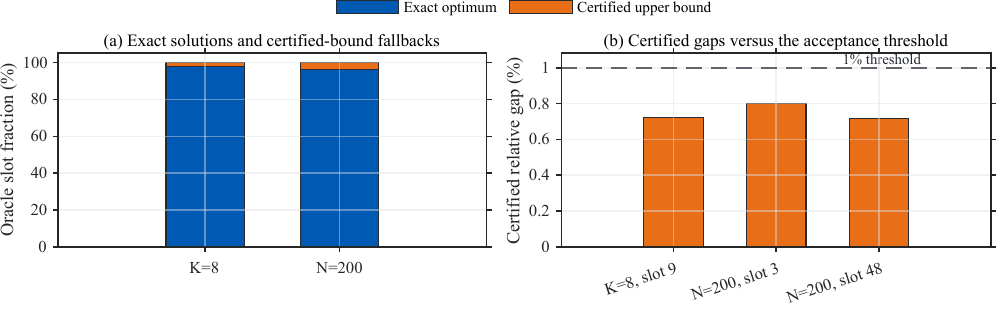}
	\caption{Exact-MILP feasibility, objective-order, and certified-bound audit for the designated hard instances.}
	\label{fig:mip_conditioning}
\end{figure*}

CAP-PD-UA reaches 99.26\% of the predicted-state MILP objective. Its simultaneous utility, coverage, and outage improvement follows from resource-specific scarcity prices and primal recovery, which redirect users before one constrained dimension invalidates an otherwise attractive link.

\subsection{Mapping Ablations}

Fig.~\ref{fig:mapping_ablation} holds traffic and association logic fixed. TL-QRM improves utility by 10.64\% and lowers outage by 5.70 points relative to Static-LUT. Target-Agnostic and Coarse-Map lose most of this gain because a single request vector cannot reflect candidate state and aggregated resources conceal accelerator, memory, and cache incompatibilities that appear only during execution.

Quantization reduces signaling by 19.84\% with less than 0.01\% utility change. The much larger Oracle-Map gap, despite 99.26\% predicted-state association optimality, identifies mapping and state uncertainty as the dominant remaining loss.

\subsection{Mapping Robustness}

Fig.~\ref{fig:sensitivity} applies common resource scales and hidden mapping errors to every mapping variant. At the highest scale, TL-QRM retains a 5.86-unit utility and 2.51-point outage advantage over Static-LUT. Coarse aggregation continues to admit requests that violate accelerator or memory compatibility, whereas target-dependent mapping rejects or redirects them before execution. The advantage therefore persists beyond a favorable reference point.

At 30\% hidden error, TL-QRM still exceeds Static-LUT by 6.04 utility units and lowers outage by 2.47 points. Online calibration reduces mapping SMAPE by 6.33\%, while its short-run capacity effect is small because few corrected errors cross a feasibility boundary.

\subsection{Association Robustness}

Fig.~\ref{fig:convergence} exposes every association method to the same scale and error grid. Under 30\% mapping error, Proposed retains a 2.24-unit utility and 1.24-point outage advantage over Price-Iter. Max-SINR, Comm-Load, and the bandits lose more because their scores omit the joint scarcity prices of the nine dimensions. At high pressure the gap narrows as the feasible set contracts for every association method, but the ordering remains unchanged.

Projected recovery and local improvement return a capacity-feasible assignment within 50 iterations. The 99.26\% MILP ratio directly measures the quality of the finite discrete solution after its feasible objective stabilizes.

\subsection{Control Sensitivity}

Fig.~\ref{fig:k_slot_delay} explains the default $K=6$ and one-second slot. Expanding $K$ from 2 to 12 raises utility by 17.5\% and lowers outage by 9.26 points, but controller time grows more than sevenfold. At $K=6$, 92\% of slots meet the deadline; gains beyond $K=10$ are small because additional visible LEOs rarely change the best feasible assignment.

Longer slots admit more executable work but age the state, making one second a practical balance. A one-second control delay reduces utility by 0.93\% and raises outage by 0.56 points. Candidate overlap and conservative estimates absorb short staleness. Correlated feeder outages and longer missing-report intervals are reserved for a physical control-plane extension.

\subsection{Runtime Scaling}

Fig.~\ref{fig:runtime} uses serial execution for interpretable timing. All slots meet the one-second deadline through 80 users and 98\% do so at 100 users, whereas only half meet it at 120 users. Thus the present serial implementation has a real-time region of about 100 users; batched candidate evaluation and parallel primal recovery can extend it.

At 120 users, CAP-PD-UA gains 1.90\% utility and lowers outage by 1.18 points relative to Price-Iter, at the cost of the nine-resource price and recovery loop. Candidate pruning, warm starts, and parallelization are therefore important under a strict deadline.

\subsection{Session Dynamics}

The 600-slot experiment in Fig.~\ref{fig:temporal} preserves session identities and resource holding, charges handovers, caps active users at 120, and drops a session after five unserved slots. Repeated per-epoch association is evaluated under persistence, with horizon planning retained as a future extension. Sustained overload produces 42.44\% admission blocking, while 81.76\% of admitted and 92.27\% of resolved sessions complete.

Accepted-session slot outage is 1.06\%, showing that admission control protects active sessions by shifting sustained overload into access blocking. This reliability--blocking tradeoff is hidden by single-slot averages and motivates duration-aware admission.

\subsection{MILP Audit}

Fig.~\ref{fig:mip_conditioning} summarizes the scaled exact-MILP audit. Row scaling removes unit imbalance among rates, cycles, bytes, and energy. All 26 hard instances return feasible references with consistent objective ordering. Three use certified upper bounds after reaching the extended limit, each with a gap below 0.8\%; all others satisfy the exact tolerance.

\FloatBarrier
\section{Conclusion}
\label{sec:conclusion}

This paper developed a mapping-aware association framework for MEO-assisted LEO networks. MGRR, SP-CCRM, and TL-QRM generate target-dependent resource vectors, while CAP-PD-UA enforces coupled resource and thermal constraints from the fused NOCC view and online calibration limits mapping bias. Across common loads, errors, and traffic seeds, utility improves by 10.64\% over Static-LUT and quantization reduces signaling by 19.84\% with negligible utility loss. Exact-MILP and system-level evaluations characterize the operating region. Future work will incorporate flight telemetry, physical control links, replicated NOCCs, and inter-LEO migration.

\bibliographystyle{IEEEtran}
\bibliography{mapping}

\end{document}